\documentclass[acmsmall]{acmart}
\usepackage{listings}
\usepackage{xcolor}
\usepackage{subfigure}
\usepackage{mathpartir}
\usepackage{syntax}
\usepackage{longtable}
\usepackage{tcolorbox}
\usepackage{tikz}
\newcommand{\circled}[1]{\tikz[baseline=(char.base)]{          \node[shape=circle,draw,inner sep=0.5pt] (char) {#1};}}

\definecolor{colorTalia}{RGB}{0,200,255}
\definecolor{colorMax}{RGB}{0,204,0}
\definecolor{colorCosmo}{RGB}{255,0,0}

\lstdefinelanguage{coq}{
    keywords={Repair, Lift, module, Module, Theorem, Proof, Record, Lemma, Definition, Abort, Qed, forall, $\forall$, Inductive, Type, Prop, Set, fun, fix, forall, Require, Import, Fixpoint, match, end, with, as, return, struct, Qed, Defined, let, Parameter, Axiom, Patch, Configure, Preprocess, Instance},
    basicstyle=\linespread{0.95}\small\ttfamily,
    keywordstyle=\color{blue},
    commentstyle=\itshape\rmfamily,
    showstringspaces=false,
    columns=flexible,
    breaklines=true,
    texcl=true,
    mathescape=true,
    tabsize=4,
    stringstyle=\color{brown},
    escapeinside={(@}{@)},
}

\lstdefinelanguage{cubicalagda}{
    keywords={data, =, $\forall$, where, eq/, squash/, _/_, [_], $\lambda$ $\equiv$},
    basicstyle=\linespread{0.95}\small\ttfamily,
    keywordstyle=\color{blue},
    commentstyle=\itshape\rmfamily,
    showstringspaces=false,
    columns=flexible,
    breaklines=true,
    texcl=true,
    mathescape=true,
    tabsize=4,
    stringstyle=\color{brown},
    escapeinside={(@}{@)},
}

\newcommand{\outline}[1]{}
\renewcommand{\outline}[1]{{\color{orange} {#1}}}

\newcommand{\trim}[1]{}
\renewcommand{\trim}[1]{{\color{blue} Can trim: {#1}}}

\setcopyright{cc}
\setcctype{by}
\acmDOI{10.1145/3763164}
\acmYear{2025}
\acmJournal{PACMPL}
\acmVolume{9}
\acmNumber{OOPSLA2}
\acmArticle{386}
\acmMonth{10}
\received{2025-03-26}
\received[accepted]{2025-08-12}

\begin{document}
\title{Proof Repair across Quotient Type Equivalences}

\author{Cosmo Viola}
\affiliation{%
  \institution{University of Illinois Urbana-Champaign}
  \country{USA}}

\author{Max Fan}
\affiliation{%
  \institution{Cornell University}
  \country{USA}}

\author{Talia Ringer}
\affiliation{%
  \institution{University of Illinois Urbana-Champaign}
  \country{USA}}


\begin{abstract}
  Proofs in proof assistants like Rocq can be brittle, breaking
  easily in response to changes. To address this, recent work introduced an algorithm and tool in Rocq to automatically repair broken proofs in response to changes that correspond to type equivalences.
  However, many changes remained out of the scope of this algorithm
  and tool---especially changes in underlying \emph{behavior}.
  We extend this proof repair algorithm so that it
  can express certain changes in behavior that were
  previously out of scope.
  We focus in particular on equivalences between 
  \emph{quotient types}---types equipped with a relation
  that describes what it means for any two elements of 
  that type to be equal.
  Quotient type equivalences can be used to express interesting
  changes in representations of mathematical structures, as well as changes in the
  implementations of data structures.

  We extend this algorithm and tool to support
  quotient type equivalences in Rocq.
  Notably, since Rocq lacks quotient types entirely,
  our extensions use Rocq's setoid machinery in place of quotients. Specifically, (1) our extension to the algorithm supports new changes corresponding to setoids, and (2) our extension to the tool supports this new class of
  changes and further automates away some of the
  new proof obligations.
  We demonstrate our extensions on proof repair
  case studies for previously unsupported changes.
  We also perform manual proof repair in Cubical Agda, a language with a univalent metatheory, which allows us to construct the first ever internal proofs of correctness for proof repair.
\end{abstract}

\begin{CCSXML}
<ccs2012>
   <concept>
       <concept_id>10011007.10011074.10011111.10011113</concept_id>
       <concept_desc>Software and its engineering~Software evolution</concept_desc>
       <concept_significance>500</concept_significance>
       </concept>
   <concept>
       <concept_id>10003752.10003790.10011740</concept_id>
       <concept_desc>Theory of computation~Type theory</concept_desc>
       <concept_significance>500</concept_significance>
       </concept>
 </ccs2012>
\end{CCSXML}

\ccsdesc[500]{Software and its engineering~Software evolution}
\ccsdesc[500]{Theory of computation~Type theory}

\keywords{Proof Repair, Cubical Agda, Rocq, Quotient Types}

\maketitle

\section{Introduction}
\label{sec:intro}
Writing formal proofs in proof assistants like Rocq\footnote{Formerly known as Coq.}, Agda, Lean, and Isabelle/HOL is a time-intensive task.
Even once written, proofs may break in the face of minor changes in the datatypes, programs,
and specifications they are about. User study data suggests that this process of writing and rewriting proofs 
is ubiquitous during proof development~\cite{ringer2019}, and that it can be challenging to deal with
even for experts.

\textit{Proof repair}~\cite{ringer2021proof} aims to simplify this process by introducing algorithms and tools that
fix formal proofs in response to breaking changes. Given an existing type $A$ and some set of functions and theorems on that type, as well a new type $B$, proof repair seeks to generate the equivalent functions and theorems defined on the type $B$. Proof repair further requires that the new functions and proofs on $B$ make no reference to the original functions and proofs on $A$, which distinguishes it from the more general \textit{proof transfer}.

Prior work introduced \textsc{Pumpkin P}i, a Rocq plugin that performs proof repair across 
changes in datatypes that can be described by type equivalences~\cite{ringer2021pldi}.
In this work, we extend proof repair to support equivalences between
\emph{quotient types} (Section~\ref{sec:setoids}).
Recent work by \citet{angiuli2021} in Cubical Agda showed that certain relations describing changes in behavior can be adjusted to equivalences between higher inductive types. One specific example presented in that work uses two queue representations. The first, one list queues, enqueue elements to the front of the list and dequeue elements from the back of the list. The second, two list queues, use a pair of lists. Elements are enqueued onto the front of the first list. When dequeuing, if the second list is empty, the first list is reversed onto the second, and then in any case the front element of the second list is removed. These types are not equivalent in any natural way, because multiple two list queues correspond to a single one list queue: for instance, the one list queue \lstinline{l} corresponds to both \lstinline{(l, [])} and \lstinline{([], rev l)}. Angiuli et al. use a higher inductive type to construct the quotient type of two list queues so that, in the resulting quotient type, \lstinline{(l, [])} and \lstinline{([], rev l)} are in the same equivalence class, which makes the types equivalent. They are then able to perform proof transfer across that equivalence.

We wish to use this same approach to implement \emph{proof repair} instead of \emph{proof transfer}. Unlike Cubical Agda, Rocq lacks quotient types entirely, so one cannot use
the original \textsc{Pumpkin P}i transformation 
as-is to support this class of changes.
To handle this, we replace quotient types by using Rocq's setoid machinery, and we replace quotient type equivalences with setoid equivalences (Section~\ref{sec:setoids}).
We then extend the proof transformation to support the 
newly generated equality proof obligations (Section~\ref{sec:approach-coq}).

By extending \textsc{Pumpkin P}i to setoids, users gain both expressive power and efficiency. Prior to this work, \textsc{Pumpkin P}i could capture a similar approach using sigma types instead of setoids. To do this, one would choose a canonical element for each equivalence class in the quotient, and then use the subtype of these elements instead of a quotient or setoid. However, that approach has severe drawbacks. In the two list queue example above, when using both quotient and setoids representations, the multiple representatives of each class allow for an amortized constant time dequeue operation. The subtype representation instead only allows for a linear time dequeue operation. 

We implement this algorithm by way of an
extension to the implementation of \textsc{Pumpkin P}i (Section~\ref{sec:automation-coq}).
Our implementation includes new automation, both to support repair across this class of changes and to
automate away some of the newly generated proof obligations corresponding to this class of changes.
We demonstrate our extended implementation on
three case studies that cannot be handled by prior proof repair work---two that are mathematical in nature and one that deals with changes in behavior (Section~\ref{sec:overview}).
\iffalse
\begin{enumerate}
    \item \textbf{We examine two representations of the integers}: (a) a standard representation, and (b) the Grothendieck group completion of the natural numbers. We repair the addition function and proofs about it (Section~\ref{sec:case1}).
    \item \textbf{We implement two variations of queues}: (a) those backed by lists and (b) those backed by \emph{pairs} of lists. These types are not equivalent, but we can construct a quotient type equivalence that describes the change and repair functions and proofs across it (Section~\ref{sec:case2}).
    \item \textbf{We move between two representations of polynomials}: (a) lists of the coefficients for each degree and (b) lists of pairs of coefficients and exponents, each pair representing a monomial (Section~\ref{sec:case3}).
\end{enumerate}
\fi
Finally, we define correctness for proof repair and construct internal proofs of correct repair in Cubical Agda (Section~\ref{sec:discussion}).
Our contributions are:

\begin{enumerate}
    \item an \textbf{extension} to the algorithm for proof repair across type equivalences that supports quotient type equivalences represented via setoids,
    \item an \textbf{implementation} of this extension in \textsc{Pumpkin P}i,
    \item new \textbf{automation} in this implementation to minimize the proof burden of setoid-specific proof obligations,
    \item a \textbf{demonstration} of the above supporting new use cases by way of three case studies, and
    \item the first \textbf{construction} of internal proofs of correct repair.
\end{enumerate}

Our code is available on Github, and we have packaged the extension, our case studies, and all the tools required to run them into an artifact \cite{artifact} for this paper.\footnote{Throughout the paper, links to specific files in the Github source have been provided where relevant. These links are given as circled numbers, like \circled{1}.}

\section{Quotients and Setoids}
\label{sec:setoids}

Our goal is to extend \textsc{Pumpkin P}i to perform repair between types equivalent up to quotients. To begin this task, we define quotient types \cite{hofmann1995extensional}:

\begin{definition} A quotient type \lstinline[mathescape=true]{A / eqA} is defined by a type \lstinline[mathescape=true]{A} and an equivalence relation \linebreak \lstinline[mathescape=true]{eqA : A $\to$ A $\to$ Prop}. \lstinline[mathescape=true]{A / eqA} has one constructor, \lstinline[mathescape=true]{[ $\cdot$ ] : A $\to$ A / eqA}, and its elements are called the \textit{equivalence classes} of the elements of \lstinline[mathescape=true]{A}. \lstinline[mathescape=true]{A / eqA} has the additional property that, for \lstinline[mathescape=true]{a, b : A, [ a ] = [ b ]} if \lstinline[mathescape=true]{eqA a b} is inhabited. Eliminating an element \lstinline[mathescape=true]{q : A / eqA} yields the underlying \lstinline[mathescape=true]{a : A} from which it was constructed. However, the user must prove that the result of the computation that is being defined is equal for any \lstinline[mathescape=true]{a, b : A} with \lstinline[mathescape=true]{[ a ] = [ b ]}. 
\end{definition} 
To give a concrete example, consider the quotient of $\mathbb{N}$ by the equivalence relation \lstinline[mathescape=true]{eq$\mathbb{N}$}, where \lstinline[mathescape=true]{eq$\mathbb{N}$ n m} is inhabited if \lstinline[mathescape=true]{n} and \lstinline[mathescape=true]{m} have the same parity. The resulting type, denoted by mathematicians as $\mathbb{N}/2$, has two elements, \lstinline[mathescape=true]{[ 0 ]} and \lstinline[mathescape=true]{[ 1 ]}; any other application of \lstinline[mathescape=true]{[ $\cdot$ ]} produces a term equal to one of these. For $\mathbb{N}/2$, eliminating \lstinline[mathescape=true]{[ 0 ]} and eliminating \lstinline[mathescape=true]{[ 2 ]} must provably result in equal terms.

Quotient constructions have historically been challenging to implement directly in constructive logic. Because of the equality property mentioned above, many type systems, including Rocq, do not support quotient types without the use of axioms. Adding quotient types can have nonobvious consequences as well. For instance, extending intensional Martin-L\"{o}f type theory with effective quotients and uniqueness of identity proofs is sufficient to prove the law of the excluded middle for types in the first universe if the type theory has at least two universes \cite{maietti1998effective}. A common alternative approach, introduced by Bishop \cite{Bishop1967} for constructive analysis and popularized in type theory by \citet{hofmann1993elimination, hofmann1995extensional}, is to use setoids. 

\begin{definition}
A setoid is the pair \lstinline{(A, eqA)}. When the equivalence relation \lstinline{eqA} is obvious from context, we may also call \lstinline{A} a setoid as shorthand.
\end{definition}

Alternative definitions exist; see \citet{JFP2003} for a survey of these definitions. Setoids are a natural alternative when actually constructing a quotient, and in turn its equivalence classes, is unfeasible. The Rocq standard library provides both definitions for setoids and substantial automation for their use. Quotients, on the other hand, see no such support. Their implementation would require the addition of axioms, which block computation and are generally avoided by Rocq users when possible. Thus, in Rocq, using setoids in place of quotients is the norm \cite{reasoningWithEqualities, PGL-045}. Furthermore, using axioms in a Rocq plugin forces users of that tool to adopt the axiom, so using axioms in plugins is doubly discouraged.

Notice that setoids have no special constructors, eliminators, or equality properties. An element of \lstinline[mathescape=true]{A} is said to be an element of the setoid, and users of the setoid should compare elements using \lstinline[mathescape=true]{eqA} instead of native equality. Because of this, rewriting is more difficult in setoids than in quotient types. Rewriting by an equality in a quotient type can be done using the equality eliminator, but rewriting cannot generally be done for arbitrary equivalence relations. 
To compensate for this, we need the notion of a proper function.
\begin{definition}
For setoids \lstinline[mathescape=true]{(A, eqA)} and \lstinline[mathescape=true]{(B, eqB)}, a function \lstinline[mathescape=true]{f : A $\to$ B} is \emph{proper} if, for all \lstinline[mathescape=true]{a1, a2 : A}, \lstinline[mathescape=true]{eqA a1 a2} implies that \lstinline[mathescape=true]{eqB (f a1)  (f a2)}. \cite{setoid, coq-morphisms} 
\end{definition} 
Using our example, $\mathbb{N}/2$, \lstinline[mathescape=true]{f : $\mathbb{N}$ $\to$ $\mathbb{N}$} is proper if inputs to \lstinline[mathescape=true]{f} of the same parity produce outputs of the same parity. The successor function would be one example of such a proper function. When a term is composed of proper functions, it becomes possible to construct proofs for rewriting by equivalence relations in much the same way we can rewrite by equality. Rocq provides automation in its standard library to support constructing these proofs. By using setoids in place of quotients, equivalence relations in place of equality, and proper functions in place of arbitrary functions, we mimic the functionality of quotients using setoids.

Our paper deals with repair across quotient type equivalences. To begin, we recall the definition of a type isomorphisms: 
\begin{definition}
An isomorphism from type \lstinline{A} to type \lstinline{B} is a function \lstinline{f : A $\to$ B} such that there exists \lstinline{g : B $\to$ A} with
  \begin{itemize}
    \item \lstinline{$\forall$ (a : A), g (f a) = a}
    \item \lstinline{$\forall$ (b : B), f (g b) = b} \cite{hottbook}
\end{itemize}
\end{definition}
Type equivalences are type isomorphisms with an additional adjoint property, which can be derived from any isomorphism. A quotient type equivalence is simply an equivalence between two quotient types. When using setoids to represent quotients, we must instead use a notion of a setoid equivalence:
\begin{definition}
A setoid equivalence between \lstinline{(A, eqA)} and \lstinline{(B, eqB)} is a function \linebreak \lstinline{f : A $\to$ B} such that there exists \lstinline{g : B $\to$ A} satisfying the following properties:
\begin{itemize}
    \item \lstinline{f} and \lstinline{g} are proper. 
    \item \lstinline{$\forall$ (a : A), eqA (g (f a))} \lstinline{a}
    \item \lstinline{$\forall$ (b : B), eqB (f (g b))} \lstinline{b} \cite{Bishop1967}
\end{itemize}
The function \lstinline{g} is said to be inverse to \lstinline{f}.
\end{definition}
Two setoids are equivalent if the quotient types they represent would be isomorphic.
One example of a nontrivial setoid equivalence between the setoids \lstinline[mathescape=true]{($\mathbb{N}$, eq$\mathbb{N}$)} and \lstinline{(Bool, =)} is given by the function \lstinline[mathescape=true]{isEven : $\mathbb{N}$ $\to$ Bool}, sending even numbers to \lstinline{true} and odd numbers to \lstinline{false}. This function has inverse \lstinline[mathescape=true]{g : Bool $\to$ $\mathbb{N}$} mapping \lstinline{true} to 0 and \lstinline{false} to 1. \lstinline{f} is proper, since \lstinline{isEven} respects parity, and \lstinline{g} is automatically proper because \lstinline{Bool} uses equality as its equivalence relation. The other conditions, \lstinline{isEven (g b)  = b} and \lstinline[mathescape=true]{eq$\mathbb{N}$ (g (isEven n)) n}, are both satisfied, and thus this indeed forms a setoid equivalence.
\section{Approach: Proof Term Transformation}
\label{sec:approach-coq}

To reiterate, our goal is to perform proof repair across quotient type equivalences. Because we are working in Rocq, which does not have native quotient types, we represent quotients using setoids. Thus, given two types \lstinline{A} and \lstinline{B}, each of which the user may choose to represent as either a type using standard equality or a setoid using some equivalence relation, we wish to turn terms defined over \lstinline{A} into terms defined over \lstinline{B} which no longer have any references to \lstinline{A}. 

To make the proof repair problem more concrete, we consider a specific example from the \textsc{Pumpkin P}i paper. In Figure~\ref{fig:binary}, we see two representations of the natural numbers. The first is a unary representation, while the second is a binary representation. Suppose that a user has begun proof development using the unary representation, implementing functions, such as addition, and theorems, such as that 0 is an identity for addition. Then, suppose that the user needs to switch to the binary representation later in development. Proof repair takes the functions and theorems defined over the unary natural numbers and turns them into functions defined over the binary natural numbers with the same behavior. These functions and theorems will make no reference to the unary natural numbers, so that the type can be deleted from the codebase entirely.

We will review how \textsc{Pumpkin P}i repairs proofs across 
type equivalences by directly transforming proof terms across those equivalences (Section~\ref{sec:transformation}).
Then, we extend \textsc{Pumpkin P}i's transformation to
support setoid equivalences (Section~\ref{sec:extended-transformation}). 

\begin{figure}
\begin{minipage}{0.39\columnwidth}
    \begin{lstlisting}[language=coq]
    Inductive nat :=
    | O : nat
    | S : nat -> nat.
    \end{lstlisting}
\end{minipage}
\hfill
\begin{minipage}{0.6\columnwidth}
    \begin{lstlisting}[language=coq]
    Inductive positive :=
    | xO : positive -> positive
    | xI : positive -> positive
    | xH : positive.
    Inductive N :=
    | N0 : N
    | Npos : positive -> N.
    \end{lstlisting}
\end{minipage}
    \caption{The natural numbers, as represented in Rocq's standard library, in unary (left) and binary (right). The type \lstinline{positive} represents all positive natural numbers. \lstinline{xH} is one, \lstinline{xO n} is appending a 0 to the right side of the binary representation of \lstinline{n}, and \lstinline{xI n} is appending a 1 to the right side of the binary representation of \lstinline{n}. Then, \lstinline{N} is either 0, as \lstinline{N0}, or a positive binary number.}
    \label{fig:binary}
\end{figure}

\subsection{\textsc{Pumpkin P}i's Transformation}
\label{sec:transformation}

\begin{figure}
    \small
    \begin{grammar}
    <i> $\in \mathbb{N}$, <v> $\in$ Vars, <s> $\in$ \{ Prop, Set, Type<i> \}
    
    <t> ::= <v> \hspace{0.06cm} | \hspace{0.06cm} <s> \hspace{0.06cm} | \hspace{0.06cm} $\Pi$ (<v> : <t>) . <t> \hspace{0.06cm} | \hspace{0.06cm} $\lambda$ (<v> : <t>) . <t> \hspace{0.06cm} | \hspace{0.06cm} <t> <t> \hspace{0.06cm} | \hspace{0.06cm} Ind (<v> : <t>)\{<t>,\ldots,<t>\} \hspace{0.06cm} \\ | \hspace{0.06cm} Constr (<i>, <t>) \hspace{0.06cm} | \hspace{0.06cm}~Elim(<t>,~<t>)\{<t>,\ldots,<t>\}
    \end{grammar}
    \caption{The grammar of CIC$_\omega$ from \textsc{Pumpkin P}i \cite{ringer2021pldi}, adapted from Timany and Jacobs \cite{timany2015CICgrammar}. 
    The terms here are, in order: variables, sorts, dependent product types, functions, applications, inductive types, constructors, and eliminators.}
    \label{fig:CIC-grammar}
\end{figure}

The \textsc{Pumpkin P}i transformation that we extend 
operates over terms in
the type theory of Rocq, the Calculus of Inductive Constructions (CIC$_\omega$)~\cite{Coquand1990}.
CIC$_\omega$ extends the Calculus of Constructions \cite{Coquand1988} with inductive types. The grammar for CIC$_\omega$ is in Figure~\ref{fig:CIC-grammar}; the type
rules are standard and omitted.

\textsc{Pumpkin P}i implements proof repair over terms in CIC$_{\omega}$ by
transforming proof terms implemented over an old type \lstinline{A} to instead be implemented over a new version of that type \lstinline{B}.
The key insight behind this transformation is that any equivalence between types \lstinline{A} and \lstinline{B} can be decomposed into separate
components that talk only about \lstinline{A} and only about \lstinline{B}~\cite{ringer2021proof}. Functions and proofs can be unified with applications of these components, reducing repair to a simple
proof term transformation replacing components that talk about \lstinline{A} with their counterparts over \lstinline{B}~\cite{ringer2021pldi}.

\textsc{Pumpkin P}i calls each such decomposed
equivalence a \textit{configuration}, comprising pairs of the form:

\begin{lstlisting}[language=coq, mathescape=true]
  ((DepConstr, DepElim), ($\iota$, $\eta$))
\end{lstlisting}
for types on both sides of the equivalence. \lstinline{DepConstr} and \lstinline{DepElim} are, respectively, constructors and eliminators for each type, termed \emph{dependent constructors} and \emph{dependent eliminators} internally by \textsc{Pumpkin P}i (even in cases where those constructors and eliminators might be non-dependently typed). The constructors must generate the elements of the type, and the eliminator must specify how to consume an element produced by the constructors. These constructors and eliminators must take the \emph{same shape}, 
even if \lstinline{A} and \lstinline{B} themselves have different shapes. Usually, then, for at least one of the types \lstinline{A} or \lstinline{B} \lstinline{DepConstr} and \lstinline{DepElim} will be distinct from the usual constructors and eliminators for that type.

Like the \textsc{Pumpkin P}i paper \cite{ringer2021pldi}, we will use repair between unary and binary natural numbers as an example. For the type of unary naturals in Figure~\ref{fig:binary}, we can choose \lstinline{depConstr} for \lstinline{nat} to be \lstinline{O} and \lstinline{S}, the constructors for \lstinline{nat}. However, to repair to the binary naturals, we need to provide dependent constructors for \lstinline{N}. These constructors \textit{must correspond across the equivalence}, so we must provide constructors with these types:
\begin{lstlisting}[language=coq]
  Definition depConstrNZero : N.
  Definition depConstrNSuc : N -> N.
\end{lstlisting}
These dependent constructors \emph{do not} share the type signatures of \lstinline{N}'s constructors, but rather are user-defined functions corresponding to \lstinline{nat}'s constructors. Specifically, the successor constructor for \lstinline{N} instead has type \lstinline{positive -> N}. Likewise, we can choose the eliminator for \lstinline{nat}, \lstinline{nat_rect}, as the \lstinline{depElim} for \lstinline{nat}, and the dependent eliminator for \lstinline{N} then must take the shape of \lstinline{nat}'s eliminator:
\begin{lstlisting}[language=coq]
  Definition depElimN : $\forall$ (P : N -> Type),
    (P depConstrNZero) -> 
    ($\forall$ n : N, P n -> P (depConstrNSuc n)) ->
    $\forall$ n : N, P n.
\end{lstlisting}
We give a concrete example of repairing addition in Figure~\ref{fig:binadd} 
~\href{https://github.com/uwplse/pumpkin-pi/blob/oopsla2025/plugin/coq/nonorn.v}{\circled{1}}.

\begin{figure*}
\centering
\begin{minipage}{0.48\textwidth}
\begin{lstlisting}[mathescape=true,language=coq]
Definition addNat (a b : nat) :=
  depElimNat
    (fun _ => nat -> nat)
    (fun b => b)
    (fun a IH b => depConstrNatSuc (IH b))
    a
    b.
\end{lstlisting}
\end{minipage}
\hfill
\begin{minipage}{0.46\textwidth}
\begin{lstlisting}[mathescape=true,language=coq]
Definition addN (a b : N) :=
  depElimN
    (fun _ => N -> N)
    (fun b => b)
    (fun a IH b => depConstrNSuc (IH b))
    a
    b.
\end{lstlisting}
\end{minipage}
\caption{Repairing the addition function from unary (left) to binary (right) natural numbers.}
\label{fig:binadd}
\end{figure*}

The fact that both dependent eliminators must have the same shape even when the underlying types do not is exactly why we need the remaining
element of the configuration: \lstinline{$\iota$}.
This gives the $\iota$-reduction rules, which specify how to reduce an application of a dependent eliminator to a dependent constructor.
When the shape is of the underlying type is the same as
the shape of the configuration components, as is true for for \lstinline{nat}, this will be definitional---the
proof assistant will handle it automatically.
However, if the inductive structure is different, as it is for \lstinline{N}, this will be a propositional equality. As an example, the $\iota$-reduction rule for \lstinline{depConstrNSuc} has the following type:
\begin{lstlisting}[language=coq]
  Definition iotaNSuc (P : N -> Type) (PO : P 0)
    (PS : forall x : N, P x -> P (depConstrNSuc x)) 
    (n : N) (Q : P (depConstrNSuc n) -> Type) :
    Q (PS n (depElimN P PO PS n)) -> Q (depElimN P PO PS (depConstrNSuc n))
\end{lstlisting}
We may also use a version of this rule in proofs which runs in the reverse direction: 
\begin{lstlisting}[language=coq]
  Definition iotaNSucRev (P : N -> Type) (PO : P 0)
    (PS : forall x : N, P x -> P (depConstrNSuc x)) 
    (n : N) (Q : P (depConstrNSuc n) -> Type) :
    Q (depElimN P PO PS (depConstrNSuc n)) -> Q (PS n (depElimN P PO PS n))
\end{lstlisting}
The last term, $\eta$, represents reducing the $\eta$-expansion of constructors applied to eliminators. However, we have yet to find an example where a nontrivial $\eta$ is actually necessary. Thus, for the purposes of this paper, $\eta$ will always be trivial and we will ignore it. \footnote{The example of nontrivial $\eta$ given in the \textsc{Pumpkin P}i paper can be rewritten to instead have nontrivial $\iota$ and trivial $\eta$ ~\href{https://github.com/uwplse/pumpkin-pi/blob/oopsla2025/plugin/coq/examples/ListToVectManual.v}{\circled{2}}, and we are not aware of any cases where this cannot be done. Further, while the type of $\iota$ is always clearly defined, what $\eta$ should be is unclear for many examples. For this reason, it's possible that $\eta$ should be dropped entirely, and a configuration should simply include \lstinline{DepConstr}, \lstinline{DepElim}, and $\iota$.} 

Once we have defined the components of the configuration, we are ready to do repair. First, the functions we wish to repair are converted to explicitly refer to the configuration terms. Sometimes this happens via manual user annotation, and sometimes this happens via custom unification machinery inside of \textsc{Pumpkin P}i that does this automatically for some classes of changes (see Section~\ref{sec:annotations}). Then, we follow the syntactic transformation outlined in Figure \ref{fig:pumpkin-pi-transformation}. We will discuss what it means for the repaired term to be \emph{correct} in Section~\ref{sec:discussion}.

\begin{figure*}
\begin{mathpar}
\mprset{flushleft}

\hfill\fbox{$\Gamma$ $\vdash$ $t$ $\Uparrow$ $t'$}\vspace{-0.3cm}\\

\inferrule[Dep-Elim]
  { \Gamma \vdash a \Uparrow b \\ \Gamma \vdash p_{a} \Uparrow p_b \\ \Gamma \vdash \vec{f_{a}}\phantom{l} \Uparrow \vec{f_{b}} }
  { \Gamma \vdash \mathrm{DepElim}(a,\ p_{a}) \vec{f_{a}} \Uparrow \mathrm{DepElim}(b,\ p_b) \vec{f_{b}} }

\inferrule[Dep-Constr]
{ \Gamma \vdash \vec{t}_{a} \Uparrow \vec{t}_{b} } %\\ TODO must we explicitly lift A to B if we want to handle parameters/indices?
{ \Gamma \vdash \mathrm{DepConstr}(j,\ A)\ \vec{t}_{a} \Uparrow \mathrm{DepConstr}(j,\ B)\ \vec{t}_{b}  }

\inferrule[Eta]
  { \\ }
  { \Gamma \vdash \mathrm{Eta}(A) \Uparrow \mathrm{Eta}(B) }

\inferrule[Iota]
  { \Gamma \vdash q_A \Uparrow q_B \\ \Gamma \vdash \vec{t_A} \Uparrow \vec{t_B} }
  { \Gamma \vdash \mathrm{Iota}(j,\ A,\ q_A)\ \vec{t_A} \Uparrow \mathrm{Iota}(j,\ B,\ q_B)\ \vec{t_B} }

\inferrule[Equivalence]
  { \\ }
  { \Gamma \vdash A\ \Uparrow B }

\inferrule[Constr]
{ \Gamma \vdash T \Uparrow T' \\ \Gamma \vdash \vec{t} \Uparrow \vec{t'} }
{ \Gamma \vdash \mathrm{Constr}(j,\ T)\ \vec{t} \Uparrow \mathrm{Constr}(j,\ T')\ \vec{t'} }

\inferrule[Ind]
  { \Gamma \vdash T \Uparrow T' \\ \Gamma \vdash \vec{C} \Uparrow \vec{C'}  }
  { \Gamma \vdash \mathrm{Ind} (\mathit{Ty} : T) \vec{C} \Uparrow \mathrm{Ind} (\mathit{Ty} : T') \vec{C'} }

%% Application
\inferrule[App]
 { \Gamma \vdash f \Uparrow f' \\ \Gamma \vdash t \Uparrow t'}
 { \Gamma \vdash f t \Uparrow f' t' }

\inferrule[Elim] % TODO wait why do we have c here when it clearly refers to the term we eliminate over? um
  { \Gamma \vdash c \Uparrow c' \\ \Gamma \vdash Q \Uparrow Q' \\ \Gamma \vdash \vec{f} \Uparrow \vec{f'}}
  { \Gamma \vdash \mathrm{Elim}(c, Q) \vec{f} \Uparrow \mathrm{Elim}(c', Q') \vec{f'}  }

% Lamda
\inferrule[Lam]
  { \Gamma \vdash t \Uparrow t' \\ \Gamma \vdash T \Uparrow T' \\ \Gamma,\ t : T \vdash b \Uparrow b' }
  {\Gamma \vdash \lambda (t : T).b \Uparrow \lambda (t' : T').b'}

% Product
\inferrule[Prod]
  { \Gamma \vdash t \Uparrow t' \\ \Gamma \vdash T \Uparrow T' \\ \Gamma,\ t : T \vdash b \Uparrow b' }
  {\Gamma \vdash \Pi (t : T).b \Uparrow \Pi (t' : T').b'}

\inferrule[Var]
  { v \in \mathrm{Vars} }
  {\Gamma \vdash v \Uparrow v}
%\inferrule[Sort]
%  { \\ }
%  {\Gamma \vdash s \Uparrow s}
\end{mathpar}
\caption{Transformation for repair across $A \simeq B$ with configuration \newline \lstinline{((DepConstr, DepElim), (Eta, Iota))}, from previous work~\cite{ringer2021proof}. Our work adapts and extends this transformation.}

    \centering
    \label{fig:pumpkin-pi-transformation}
    % \caption{The proof repair transformation. \outline{copy from pumpkin pi paper}}
\end{figure*}

\subsection{Extended Transformation}
\label{sec:extended-transformation}

\begin{figure*}
    \begin{mathpar}
\mprset{flushleft}
\small
\hfill\fbox{$\Gamma$ $\Uparrow$ $\Gamma'$}\vspace{-0.3cm}\\

\inferrule[LiftEmpty]
  { \\ }
  { () \Uparrow () }

\inferrule[LiftCons]
  { \Gamma \Uparrow \Gamma' \\ \Gamma \vdash x \Uparrow x' \\ \Gamma \vdash X \Uparrow X' }
  { (\Gamma, x : X) 
  \Uparrow (\Gamma', x' : X') }
\end{mathpar}
\begin{mathpar}
\mprset{flushleft}
\small
\hfill\fbox{$\Gamma$ $\vdash$ $t$ $\Uparrow$ $t'$}\vspace{-0.3cm}\\
\inferrule[EquivApp]
  { \Gamma \vdash A \Uparrow B }
  { \Gamma \vdash \mathrm{\equiv_A}\ \Uparrow\ \equiv_B }

\inferrule[ReflexivityApp]
  { \Gamma \vdash A \Uparrow B }
  { \Gamma \vdash \mathtt{reflexivity}(\equiv_A) \Uparrow \mathtt{reflexivity}(\equiv_B) }
\\
\inferrule[EqRewrite]
  { \phantom{\vdash x ,} \Gamma \Uparrow \Gamma' \\\\ \Gamma \vdash A \Uparrow B \\ \phantom{,} \Gamma \vdash x \Uparrow x' \\ \Gamma \vdash P \Uparrow P' \\\\ \Gamma \vdash f \Uparrow f' \\ \Gamma \vdash y \Uparrow y' \\ \Gamma \vdash e \Uparrow e' \\ }
  { \Gamma \vdash \mathtt{@eq\_rect}(A, x, P, f, y, e) \Uparrow \\\\ 
    \phantom{\Gamma} \lBrack \mathrm{LiftRewrite}_{\Gamma'}(B, x', P', f', y', e') \rBrack }

\inferrule[SetoidRewrite]
  { \phantom{\vdash x i} \Gamma \Uparrow \Gamma' \\\\ \Gamma \vdash A \Uparrow B \\ \phantom{,} \Gamma \vdash x \Uparrow x' \\ \Gamma \vdash y \Uparrow y' \\\\ \Gamma \vdash e \phantom{,} \Uparrow e' \\ \Gamma \vdash g \Uparrow g' \\ \phantom{,} \Gamma \vdash t \Uparrow t' }
  { \Gamma \vdash \mathtt{StartRewrite}(A, x, y, e, g, t) \Uparrow \\\\
    \phantom{\Gamma} \lBrack \mathrm{LiftSetoidRewrite}_{\Gamma'}(B, x', y', e', g', t') \rBrack}
    \end{mathpar}
    \caption{The additional rules needed for repairing across setoid equivalences. There are two mutually defined judgments: one to repair environments (top) and one to repair terms (bottom).}
    \label{fig:quotient-transform}
\end{figure*}

We now extend this transformation to work for setoid
equivalences as well.
To do this, we adapt our transformation to handle the changes in how equality works (Figure~\ref{fig:quotient-transform}). We handle equivalence and its proofs in three cases: its type, its construction by reflexivity, and its elimination by rewriting.

\paragraph{Types} To define an instance of a repair transformation using setoids, the user must provide custom equivalence relations for any type they wish to consider as a setoid. This is done by providing three terms: 
\begin{enumerate}
    \item the type \lstinline{C : Type}
    \item a binary relation \lstinline{$\equiv_C$ : C -> C -> Prop}
    \item a proof that \lstinline{$\equiv_C$} is an equivalence relation
\end{enumerate}
If the user does not supply these terms for some type, then $\equiv_C$ is be assumed to be equality. Then, if a type \lstinline{C} repairs to \lstinline{D}, all occurrences of \lstinline{$\equiv_C$} repair to \lstinline{$\equiv_D$} (by \textsc{EquivApp}).

\paragraph{Construction by Reflexivity} 
The next rule, \textsc{ReflexivityApp}, deals with constructions of equivalence relations
by reflexivity. The proof supplied by the user that \lstinline{$\equiv_C$} is an equivalence relation contains
a term for reflexivity, with type \lstinline{$\forall$ c : C, $\equiv_C$ c c}.
We denote this term \lstinline{reflexivity($\equiv_C$)}.
If $\equiv_C$ is \lstinline{@eq C}, this is simply \lstinline{eq_refl C}.
In any case, each \lstinline{reflexivity($\equiv_C$)} repairs to \lstinline{reflexivity($\equiv_D$)} (by \textsc{ReflexivityApp}).

\paragraph{Elimination by Rewriting}
For rewriting, we split into two cases depending on if \lstinline{$\equiv_C$} is \lstinline{@eq C} or some other equivalence relation.
When \lstinline{$\equiv_C$} is \lstinline{@eq C},
then the eliminators (like \lstinline{@eq_rect} in Rocq) define term rewrites, with \lstinline{P} defining where in the term rewrites take place.
Our equivalence relations do not support arbitrary term rewrites, however, so we cannot directly translate this term. Instead, we assume we have an oracle $\lBrack - \rBrack$ which can prove that a given rewrite, denoted LiftRewrite$_\Gamma$(\lstinline{D}, \lstinline{x}, \lstinline{P}, \lstinline{px}, \lstinline{y}, \lstinline{H}), can be performed. In Section~\ref{sec:automation-coq}, we will discuss how we implement this oracle without any additional axioms using Rocq's built in setoid rewriting automation. This oracle requires access to the environment $\Gamma$ so that the oracle can refer to the repaired terms when discovering the rewrite proof. The rules \textsc{LiftEmpty} and \textsc{LiftCons} describe how the environment is transformed. Then, applications of \lstinline{@eq_rect} are replaced with the proof produced by the oracle (by \textsc{EqRewrite}). We show an example of repairing such a rewrite in Figure~\ref{fig:repair-rewrite-eqrefl} ~\href{https://github.com/uwplse/pumpkin-pi/blob/oopsla2025/plugin/coq/ToSetoidTest.v}{\circled{3}}.

\begin{figure*}
\centering
\begin{minipage}{0.46\textwidth}
\begin{lstlisting}[mathescape=true,language=coq]
Inductive unit :=
  | tt.
  
Theorem rewrite_example : 
  forall (x : unit), eq x tt -> eq x tt.
Proof.
  intros x H.
  rewrite H.
  reflexivity.
Qed.

fun (x : unit) (H : @eq unit x tt) =>
  @eq_ind_r unit tt 
    (fun x0 : unit => @eq unit x0 tt)
    (reflexivity($ $@eq unit) tt) x H
\end{lstlisting}
\end{minipage}
\hfill
\begin{minipage}{0.49\textwidth}  
\begin{lstlisting}[mathescape=true,language=coq]
Inductive unit_two :=
  | one
  | two.

Definition eq_unit_two (u1 u2 : unit_two) : Prop := True.

fun (x : unit_two) 
  (H : eq_unit_two x one) =>
  [[ LiftRewrite$_{\Gamma'}$ 
    (unit_two, one, 
    (fun x0 : unit_two => 
      eq_unit_two x0 one), 
    (reflexivity(eq_unit_two) one), x, H)
  ]] : forall x : unit_two, 
         eq_unit_two x one -> 
         eq_unit_two x one
\end{lstlisting}
\end{minipage}
\caption{A demonstration of repairing a rewrite in a proof. The left side is over the source type, which is simply the type with one element. We provide a sample Rocq tactic script which performs a rewrite, and below it is the term produced. The right side shows the repaired term over the target setoid, which is the type with two elements which are equal under the equivalence relation. 
}
\label{fig:repair-rewrite-eqrefl}
\end{figure*}

If \lstinline{$\equiv_C$} is some other equivalence relation,
the user can still perform rewrites, except those rewrites behave differently.
We denote such a rewrite using \lstinline{StartRewrite(C, x, y, e, g, t)}. Here, \lstinline{x y : C}, \lstinline{e : $\equiv_C$ x y}, \lstinline{g} is the type the user is rewriting, and \lstinline{t} is the term the rewrite is applied to. This rewrite replaces every instance of \lstinline{x} in the type of \lstinline{t} with \lstinline{y}. The \textsc{SetoidRewrite} rule assumes that our oracle $\lBrack - \rBrack$ can perform a rewrite using the repaired data on the repaired term, which we denote by LiftSetoidRewrite$_{\Gamma}$(\lstinline{D}, \lstinline{x}, \lstinline{y}, \lstinline{e}, \lstinline{g} \lstinline{t}). Then, instances of \lstinline{StartRewrite(C, x, y, e, g, t)} repair to $\lBrack$LiftSetoidRewrite$_{\Gamma}$(\lstinline{D}, \lstinline{x}, \lstinline{y}, \lstinline{e}, \lstinline{g} \lstinline{t})$\rBrack$ (by \textsc{SetoidRewrite}).

Importantly, this extension to \textsc{Pumpkin P}i can only affect the trusted computing base of the proof system if the rewrite oracle has trust assumptions. Thus, by enforcing that the implementation of the oracle introduces no such assumptions, using this extension poses no risk of a soundness failure beyond that of the base type theory without any extensions.

\section{Implementation}
\label{sec:automation-coq}

We implement this extension to the transformation by
extending the \textsc{Pumpkin P}i Rocq plugin (Section~\ref{sec:extending}).
Consistently with the original \textsc{Pumpkin P}i, 
our implementation relies on some user annotations (Section~\ref{sec:annotations}),
and places some other restrictions on the format terms can take (Section~\ref{sec:restrictions}).
Our implementation includes custom automation to dispatch
properness proofs specific to setoids (Section~\ref{sec:proper}).

\subsection{Extending \textsc{Pumpkin P}i}
\label{sec:extending}

We extend the \textsc{Pumpkin P}i Rocq plugin directly. Plugins  
are a method of adding functionality to Rocq. They are written in OCaml and can interact directly with the Rocq internal codebase.
Plugins can directly transform and produce terms;
all terms that plugins produce are checked by Rocq's type checker, and so cannot be ill typed.

\textsc{Pumpkin P}i has various classes of proof repair transformations across type equivalences for which it has specialized automation. We add an additional class, termed setoid lifting, to support our extended transformation. This class mostly reuses the existing transformation, but implements the new rules from Figure~\ref{fig:quotient-transform}. Once the transformation rules were determined, the implementation was made straightforward thanks to \textsc{Pumpkin P}i's extensibility; in all, our extension adds 1659 lines of code, 510 of which are dedicated to properness proof generation implemented in \lstinline{propergen.ml}. The core extensions to the transformation can be found in \lstinline{lift.ml},
~\href{https://github.com/uwplse/pumpkin-pi/blob/oopsla2025/plugin/src/automation/lift/lift.ml}{\circled{4}}
\lstinline{liftconfig.ml},
~\href{https://github.com/uwplse/pumpkin-pi/blob/oopsla2025/plugin/src/automation/lift/liftconfig.ml}{\circled{5}}
and \lstinline{liftrules.ml}
~\href{https://github.com/uwplse/pumpkin-pi/blob/oopsla2025/plugin/src/automation/lift/liftrules.ml}{\circled{6}} 
in the artifact.

In our extended configuration, 
the proof that $\equiv_C$ is an equivalence relation
takes the form of instances of type classes \lstinline{Equivalence $\equiv_C$}. This makes it possible to use Rocq's setoid automation to implement the oracle that produces proofs of rewrites described in Section~\ref{sec:extended-transformation} with no axioms or other extensions to the trusted computing base. 
Rocq has a tactic, called \lstinline{setoid_rewrite}, which attempts to perform rewriting by an equivalence relation. However, because rewriting arbitrary terms between equivalent elements is not possible, we must \textit{prove} for each function we define that the function is proper, as defined in Section~\ref{sec:setoids}, if we wish to rewrite under applications of that function. In Rocq, we prove this by creating an instance of the \lstinline{Proper} type class. The \lstinline{setoid_rewrite} tactic uses the \lstinline{Proper} and \lstinline{Equivalence} type class instances to search for proofs of rewrites, and thus we can use it as our oracle. 

\subsection{User Annotations}
\label{sec:annotations}

To perform repair, \textsc{Pumpkin P}i requires that users annotate their proofs explicitly with components of the configuration. These annotations are required to identify parts of the configuration, thereby decoupling the undecidable part of proof repair (configuration inference) from the decidable part (the proof term transformation itself). We inherit this requirement for our extension, and thus both the original configuration and to our extension to equivalences must be annotated. However, we have implemented automation into \textsc{Pumpkin P}i that generates many of the needed annotations for equivalence relations automatically.

Specifically, if no relation is provided for a given type,
our extension defaults to strict equality. Then, we can automatically infer annotations corresponding to applications of equality \lstinline{@eq C}, as well as applications of reflexivity \lstinline{@eq_refl C}.
The same holds for rewrites that fully apply any of Rocq's equality eliminators 
~\href{https://github.com/uwplse/pumpkin-pi/blob/oopsla2025/plugin/src/automation/lift/liftconfig.ml}{\circled{5}}.
However, this annotation inference cannot deal with
unapplied instances of \lstinline{@eq} and \lstinline{@eq_refl} that may later be specialized
to \lstinline{C}, or equality eliminators that are not fully applied. Such terms are considered improperly annotated.

When an equivalence relation is provided for a type, we provide a custom tactic \lstinline{rewrite_annotate} which the user can use in place of \lstinline{rewrite} in proofs, which automatically performs annotation while rewriting. This inserts a custom annotation term \lstinline{START_REWRITE} before applying the rewrite, which our extension looks for to identify and repair rewrites. This constant takes two non-implicit arguments: the proof of equivalence by which the user is rewriting, and the type they are rewriting 
~\href{https://github.com/uwplse/pumpkin-pi/blob/oopsla2025/plugin/theories/SetoidDefs.v}{\circled{7}}.

\begin{figure*}
\begin{minipage}{0.38\textwidth}
\begin{lstlisting}[mathescape=true, language=coq]
Theorem depRec (C : Type)
  (posP : $\forall$ (n : nat), C)
  (negSucP : $\forall$ (n : nat), C)
  (z : GZ) : C.
$\newline$
\end{lstlisting}
\hfill
\end{minipage}
\begin{minipage}{0.61\textwidth}
\begin{lstlisting}[mathescape=true, language=coq]
Theorem depElimProp (P : GZ -> Prop)
  `(p : Proper (GZ -> Prop) (eq_GZ ==> iff) P)
  (posP : $\forall$ (n : nat), P (depConstrPos n))
  (negSucP : $\forall$ (n : nat), P (depConstrNegSuc n))
  (z : GZ) : P z.
\end{lstlisting}
\end{minipage}
\caption{The types of the two eliminators we use in one of our case studies. The left has non-dependently typed output, but can eliminate into \lstinline{Type}, while the right has dependently typed output but only eliminates into \lstinline{Prop}. The right eliminator also requires a proof that the motive is proper as a function from the setoid (\lstinline{GZ}, \lstinline{eq_GZ}) to the setoid (\lstinline{Prop}, \lstinline{iff}).}
\label{fig:elim-examples}
\end{figure*}

\subsection{Restrictions}
\label{sec:restrictions}

\begin{figure*}
\begin{lstlisting}[mathescape=true, language=coq]
Definition respectful_hetero
  (A B : Type)
  (C : A -> Type) (D : B -> Type)
  (R : A -> B -> Prop)
  (R' : forall (x : A) (y : B), C x -> D y -> Prop) :
    (forall x : A, C x) -> (forall x : B, D x) -> Prop :=
    fun f g => forall x y, R x y -> R' x y (f x) (g y).
\end{lstlisting}
\caption{An equivalence relation from the Rocq standard library on dependently typed functions stating that the functions are respectful. \lstinline{R} relates elements of their domains, and \lstinline{R'} relates elements of their codomains. Notice that because the functions are dependently typed, \lstinline{R'} is a family of equivalence relations, one for each pair of elements \lstinline{x : A, y : B}. Specializing \lstinline{respectful_hetero} to non-dependently typed functions is used to define the notion of a proper function we use in this work.}
\label{fig:respectful-hetero}
\end{figure*}

\textsc{Pumpkin P}i does not directly repair terms involving pattern matching and recursion, instead requiring that proof terms it repairs are written using induction principles. As a result, our extension to \textsc{Pumpkin P}i also has this limitation. However, \textsc{Pumpkin P}i includes some automation to transform pattern matching and recursion in proof terms into induction, which is likewise bundled in our extension.
We inherit \textsc{Pumpkin P}i's proof term decompiler, which makes it possible to get simple tactic proofs from repaired proof terms. 

While the original proof repair work had a single \lstinline{depElim} term on each side of the configuration, we have multiple eliminators. One eliminates into the sort \lstinline{Type}, but is purely nondependent. The other is dependent, but only eliminates into the sort \lstinline{Prop}, and requires that the motive of the eliminator be proven to be proper, considering \lstinline{Prop} as a setoid with \lstinline{iff}. The types of two of these eliminators for one of our case studies are in Figure~\ref{fig:elim-examples}. We do this because Rocq's setoid automation does not work for a dependently typed notion of a proper function. Specifically, the term \lstinline{respectful_hetero}, given in Figure~\ref{fig:respectful-hetero}, exists in the standard library \cite{coq-morphisms} and could form the basis for such automation, but no such automation has been implemented in the standard library. Thus, we cannot use the setoid automation to perform rewrites on applications of functions we define with a dependently typed eliminator. For our \lstinline{Prop}-sorted eliminator, this loss means that users cannot automatically perform rewrites on the \textit{proofs} of propositions. In Rocq, \lstinline{Prop} is frequently treated as effectively proof irrelevant, so this loss is more acceptable. 

In addition, our use of Rocq's setoid automation is facilitated through the use of Rocq's rewrite tactics. However, these tactics do not allow a rewrite which does not change the goal type, and so neither does our automation for repairing setoid rewrites.
Furthermore, the setoid rewrite automation does not allow specifying a motive \lstinline{P}. To perform setoid rewrites with a motive, a user trying to prove \lstinline{P x -> P y} can perform a substitution \lstinline{P}[\lstinline{z}/\lstinline{y}], where \lstinline{z} is free in \lstinline{P}, and then define \lstinline{Q := fun z => P}[\lstinline{z}/\lstinline{y}]. Then, for another fresh variable \lstinline{w}, the user can perform a \lstinline{setoid_rewrite} to prove \lstinline{H : $\forall$ (w : B), Q w x -> Q w y}, and recover the desired rewrite proof as \lstinline{H y}. Our extension uses this methodology when repairing rewrites with a motive to setoids 
~\href{https://github.com/uwplse/pumpkin-pi/blob/oopsla2025/plugin/src/automation/lift/lift.ml}{\circled{4}}.

\subsection{Automating Properness Proofs}
\label{sec:proper}

To perform rewrites on a term using Rocq's setoid automation, it is necessary to prove that the functions in that term are proper.
Thus, when repairing terms, it is potentially necessary to have properness proofs for every previously repaired function. 
We implement automation that helps prove many functions to be proper automatically. Our automation must fail in some cases, though, since proving a general function is proper is undecidable. To prove this, for any proposition \lstinline{P : Prop}, define \lstinline{f : bool -> Prop}, \lstinline{f b = if b then True else P}. Generating a proof that \lstinline{f} is proper, using \lstinline{bool} as a setoid relating \lstinline{true} and \lstinline{false}, and \lstinline{Prop} as a setoid with the relation \lstinline{iff}, is equivalent to proving \lstinline{iff P True}, which is undecidable. 
\iffalse
\begin{lstlisting}
  f : A -> B -> ... -> Z
\end{lstlisting}
is a composition of proper functions, our automation attempts to prove the goal:

\begin{lstlisting}
  Proper (eqA ==> eqB ==> ... ==> eqZ) f    
\end{lstlisting}
\fi

Presently, our automation constructs properness proofs automatically in two practical cases. First, when \lstinline{f} is a composition of proper functions, we introduce hypotheses stating that all arguments are equivalent, and then rewrite by these hypotheses. This is the approach taken by Rocq's \newline \lstinline{solve_proper} tactic, though we modify it slightly. Rocq's \lstinline{solve_proper} fails if any of its inputs do not appear in the body of the function because the rewrite for that argument will fail. To avoid this, we use the \lstinline{try} tactical when rewriting. We also try both Rocq's \lstinline{rewrite} and \lstinline{setoid_rewrite} tactics, while \lstinline{solve_proper} only runs \lstinline{setoid_rewrite}. While \lstinline{setoid_rewrite} supports rewriting under binders in some instances, \lstinline{rewrite} succeeds in some instances where \lstinline{setoid_rewrite} fails.

Second, suppose that \lstinline{f} is an application of the eliminator of some inductive type with a constant motive.
Then, if all of the inductive case arguments provided are proper functions, we can prove that the eliminator is a proper function from its base cases to its output. To make this concrete, we will consider an example. Let (\lstinline{C}, \lstinline{eqC}) be a setoid, fix some \lstinline{P = fun _ => C}, and consider:

\begin{lstlisting}[language=coq, mathescape=true]
  nat_rect P : $\forall$ (po : C) (ps : forall (n : nat) (pn : C), C) (n : nat), C    
\end{lstlisting}
Then, we can automatically prove that:

\begin{lstlisting}[language=coq, mathescape=true]
  Proper (eq ==> eqC ==> eqC) ps -> 
    $\forall$ (n : nat), Proper (eqC ==> eqC) (fun po => nat_rect P po ps n)    
\end{lstlisting}
by induction on \lstinline{n}. The base case holds by assumption via the definition of \lstinline{Proper}, and the inductive case holds by \lstinline{ps} being proper.

This can be done in general for inductive types. Thus, our automation checks if \lstinline{f} is an eliminator at the top level, and if so generates and tries to prove such a proper goal. If it succeeds, the automation attempts to show that the inductive cases (in the above example, \lstinline{ps}) are proper using the rewriting strategy described above 
~\href{https://github.com/uwplse/pumpkin-pi/blob/oopsla2025/plugin/src/automation/lift/propergen.ml}{\circled{8}}.
In our case studies, this is necessary for automatically solving some of the generated proper goals
~\href{https://github.com/uwplse/pumpkin-pi/blob/oopsla2025/plugin/case-studies/grothendieck_int_equivalence_repair_tool.v}{\circled{9}}.

\begin{figure}
\begin{minipage}{0.35\textwidth}
\begin{lstlisting}[mathescape=true,language=coq]
Inductive Z : Set :=
| pos : nat -> Z
| negsuc : nat -> Z.
\end{lstlisting}
\end{minipage}
\begin{minipage}{0.60\textwidth}
\begin{lstlisting}[mathescape=true,language=coq]
Definition GZ := nat * nat.
Definition eq_GZ (z1 z2 : GZ) := match z1, z2 with
| (a1, a2), (b1, b2) => a1 + b2 = a2 + b1
end.
\end{lstlisting}
\end{minipage}
\caption{The types of our integer representations. We provide an instance of \lstinline{Equivalence eq_GZ} for the case study in the artifact.}
\label{fig:int-types}
\end{figure}

\section{Case Studies}
\label{sec:overview}

We use our extended version of \textsc{Pumpkin P}i
to automatically repair proofs on three case studies that use quotient type equivalences.
First, we conduct repair between two representations of the integers (Section~\ref{sec:case1}). Second, we study two common implementations of the queue data structure and how we can repair from one to the other (Section~\ref{sec:case2}).
Third, we repair between dense and sparse representations of polynomials with natural number coefficients (Section~\ref{sec:case3}). In each of these case studies, the type to which we repair functions and theorems has a structure which enables efficient implementations of key operations. Our extension makes it possible to take advantage of this structure while using the repaired theorems, something impossible with prior versions of \textsc{Pumpkin P}i. 
All of the case study examples can be found in more detail in the artifact.

\subsection{Adding, Fast and Slow}
\label{sec:case1}

Our first case study is a usage of quotients that is foundational in mathematics. We consider a change in the type representing integers from the inductive type found in many standard libraries to the standard quotient based representation. We repair addition and proofs about addition from one representation to the other. Finally, we recover the repaired proofs for a more efficient version of addition over the repaired type. 
Our Rocq implementation of this case study can be found in the artifact 
~\href{https://github.com/uwplse/pumpkin-pi/blob/oopsla2025/plugin/case-studies/grothendieck_int_equivalence_repair_tool.v}{\circled{9}}.

\paragraph{Types \& Configuration}

Our first representation, \lstinline{Z}, is based on the default implementation of the integers in Cubical Agda: two copies
of $\mathbb{N}$ glued back-to-back.
We will repair functions and proofs about \lstinline{Z} to use a representation
frequently used as a definition in mathematics:
viewing the integers as elements of $\mathbb{N}\times\mathbb{N}/\sim$, where $(x_1, x_2)\sim (y_1, y_2) \iff x_1 + y_2 = x_2 + y_1$.
We call the resulting type \lstinline{GZ}\footnote{In reference to Grothendieck, as this is the Grothendieck group of the natural numbers}, and the equivalence relation \lstinline{eq_GZ}. The definitions are in Figure~\ref{fig:int-types}.
\lstinline{Z} and \lstinline{GZ} are setoid equivalent using the definition from Section~\ref{sec:setoids},
by mapping \lstinline{pos n} to \lstinline{(n, 0)} and \lstinline{negsuc n} to \lstinline{(0, S n)}.

We next decompose our isomorphism into a repair configuration consisting of dependent constructors, dependent eliminators, and $\iota$-reduction rules for both types.
The configuration differs from those found in the original \textsc{Pumpkin P}i examples in that there are two eliminators (Figure~\ref{fig:elim-examples} from Section~\ref{sec:restrictions}): \lstinline{depRec} for eliminating into nondependent types, and \lstinline{depElimProp} for eliminating into dependent types that reside in \lstinline{Prop}. Furthermore, \lstinline{depElimProp} on \lstinline{GZ} has an extra proof obligation: the motive \lstinline {P : GZ -> Prop} must be a proper function, where the sort \lstinline{Prop} is viewed as the setoid (\lstinline{Prop}, \lstinline{iff}). While in theory each of these eliminators need their own set of $\iota$-reduction theorems,
we provide them solely for \lstinline{depRec},
since needing $\iota$ for \lstinline{depElimProp} is not common and does not show up in our case study. The full configuration can be found in the artifact.

\begin{figure*}
\centering
\begin{minipage}{0.45\textwidth}
\begin{lstlisting}[mathescape=true,language=coq]
Definition addZ (z1 z2 : Z) : Z :=
  depRecZ Z
    (fun (p : nat) => add_posZ z1 p)
    (fun (p : nat) => add_negsucZ z1 p)
    z2.
\end{lstlisting}
\end{minipage}
\hfill
\begin{minipage}{0.45\textwidth}
\begin{lstlisting}[mathescape=true,language=coq]
Definition addGZ (z1 z2 : GZ) : GZ := 
  depRecGZ GZ 
    (fun p : nat => add_posGZ z1 p) 
    (fun p : nat => add_negsucGZ z1 p) 
    z2.
\end{lstlisting}
\end{minipage}
\caption{The annotated definition of addition on \lstinline{Z} (left, adapted from the Cubical Agda standard library), and the repaired function defined over \lstinline{GZ} (right). We omit the definitions of \lstinline{add_posZ} and \lstinline{add_negsecZ}, which are also automatically repaired.}
\label{fig:z-standard-add-repair}
\end{figure*}

\paragraph{Function Repair}
Next, we repair functions automatically using our extension
of \textsc{Pumpkin P}i.
For example, we repair addition from \lstinline{Z} (Figure~\ref{fig:z-standard-add-repair}, left) to \lstinline{GZ} (Figure~\ref{fig:z-standard-add-repair}, right) by running the following command:

\begin{lstlisting}[language=coq]
  Lift Z GZ in addZ as addGZ.
\end{lstlisting}
Note that the call to \lstinline{depRecZ} is directly replaced with one to \lstinline{depRecGZ}, and the functions \lstinline{add_posZ} and \lstinline{add_negsucZ} are replaced with their repaired analogues. We also repair the successor and predecessor functions.

Our extension of \textsc{Pumpkin P}i also automatically generates proofs that the repaired functions are proper.
First, the user must prove that \lstinline{depRec} is proper. 
Our extension uses this to generate
properness proofs for functions applying \lstinline{depRec}. For example, it generates the proof that addition is proper:

\begin{lstlisting}[language=coq, mathescape=true]
  addGZ_proper : Proper (eq_GZ ==> eq_GZ ==> eq_GZ) addGZ.
\end{lstlisting}

\begin{figure}
\centering
\begin{minipage}{0.45\textwidth}
\begin{lstlisting}[mathescape=true,language=coq]
Theorem add0LZ :  $\forall$ z : Z, 
  z = addZ (depConstrZPos 0) z.
\end{lstlisting}
\end{minipage}
\hfill
\begin{minipage}{0.45\textwidth}
\begin{lstlisting}[mathescape=true,language=coq]
Theorem add0LGZ : $\forall$ z : GZ, 
  eq_GZ z (addGZ (depConstrGZPos 0) z).
\end{lstlisting}
\end{minipage}
\caption{An addition identity whose proof we repaired automatically.}
\label{fig:add0LZ}
\end{figure}

\paragraph{Proof Repair}
We automatically repair the proof \lstinline{add0LZ},
which shows that $0$ is a left identity for addition.
Figure~\ref{fig:add0LZ} shows the old and new theorem types. Note that, in the repaired theorem type, equality has been automatically replaced with an equivalence relation on the type, reflecting that the repaired theorem is about setoid equality instead of \lstinline{eq}. In addition, we repair a proof that $0$ is a right identity for addition.
The original and repaired proofs can be found in the artifact.

For now, there is a bit of extra work related to proper
proof generation when our proofs apply \lstinline{depElimPropGZ} (as \lstinline{add0LZ} does).
In particular, while our automation generates properness proofs
for some of the motives passed to \lstinline{depElimPropGZ}, there is not yet a way to automatically supply those proofs to \textsc{Pumpkin P}i so that it uses them when repairing applications of \lstinline{depElimPropZ}. For now, we define a constant corresponding to the motive, which we then separately repair:

\begin{lstlisting}[language=coq, mathescape=true]
  Lift Z GZ in add0LMotiveZ as add0LMotiveGZ.
\end{lstlisting}

The proof that this motive is proper is automatically generated. We then reconfigure \textsc{Pumpkin P}i to use these applications of \lstinline{depElimPropZ} and \lstinline{depElimPropGZ} for its eliminators:

\begin{lstlisting}[language=coq, mathescape=true]
  Definition appliedDepElimPropZ := 
    depElimPropZ add0LMotiveZ.
  Definition appliedDepElimPropGZ :=
    depElimPropGZ add0LMotiveGZ add0LMotive_proper.
\end{lstlisting}
We use \lstinline {appliedDepElimPropZ} in our proof of \lstinline{add0LZ}, and can repair the term. Presently, an implementation bug only surfacing in this case study forces us to repair the arguments to \newline \lstinline{appliedDepElimPropZ} before reconfiguring.
Future versions can avoid the need for this workaround by automatically supplying the necessary properness proofs to \lstinline{depElimProp} terms. 

\paragraph{Further Steps}
As is, our repaired addition function is inefficient.
It uses the repaired eliminator for \lstinline{GZ}, which inherits the inductive structure of 
the eliminator for \lstinline{Z}.
This repaired eliminator is slow, as 
it internally computes a canonical representative of the equivalence class of the given element.

\begin{figure}
\begin{lstlisting}[mathescape=true, language=coq]
Definition fastAddGZ (a b : GZ) := match b with
  | (b1, b2) => match a with
    | (a1, a2) => (a1 + b1, a2 + b2)
  end
end.
\end{lstlisting}
\caption{Our fast addition function on the repaired integers. The direct use of pattern matching is acceptable because this function is neither to be repaired nor a product of repair.}
\label{fig:fastadd}
\end{figure}
We adapt our repaired proofs to use the more efficient addition function defined in Figure~\ref{fig:fastadd}.
Consistently with prior work in \textsc{Pumpkin P}i,
to move between slow and fast implementations we take an ad hoc approach.
First, we prove that both implementations of addition are pointwise equivalent in our setoid, producing a term \lstinline{addEqualFastAdd}.
Then, we rewrite across this equality to get from proofs of repaired theorems defined over slow addition
to proofs of corresponding theorems defined over fast addition.
For example, we use this methodology to translate the proof of \lstinline{add0LGZ} into the proof of theorem: 
\begin{lstlisting}[language=coq]
  Theorem fastAdd0LGZ : $\forall$ (z : GZ),
    eq_GZ z (fastAddGZ (depConstrGZPos 0) z).
\end{lstlisting}
using only one rewrite by \lstinline{addEqualFastAdd}.

\iffalse
If applications of \lstinline{addGZ} were inside opaque terms, however, we may not be able to view all the applications of \lstinline{addGZ}, and thus could not rewrite like this. Higher order functions are one potential source of this trouble; for instance, if the definition of \lstinline{List.map} is opaque, and we pass \lstinline{addGZ} as the function to map over a list, we would not be able to access the application sites of \lstinline{addGZ} and thus could not rewrite by \lstinline{addEqualFastAdd}. If we knew that \lstinline{fastAddGZ = addGZ}, we could perform the rewrite anyway, which is one advantage we would get in a type theory with functional extensionality like Cubical Agda (see Section~\ref{sec:disc-fast}).
\fi

\begin{figure}
\centering
\begin{minipage}{0.35\textwidth}
\begin{lstlisting}[mathescape=true, language=coq]
Definition OLQ := list A.
\end{lstlisting}
\end{minipage}
\begin{minipage}{0.55\textwidth}
\begin{lstlisting}[mathescape=true, language=coq]
Definition TLQ := list A * list A.
Definition insOrder (q : TLQ) := match q with
| (l1, l2) => l1 ++ rev l2
end.
Definition eq_queue (q1 q2 : TLQ) := 
  insOrder q1 = insOrder q2.
\end{lstlisting}
\end{minipage}
\caption{One list queues and two list queues. We provide an instance of \lstinline{Equivalence eq_queue} in the artifact.}
\label{fig:queue-types}
\end{figure}

\begin{figure}
\centering
\begin{minipage}{0.485\textwidth}
\begin{lstlisting}[mathescape=true, language=coq]
Definition enqueueOLQ (a : A) (q : OLQ) : OLQ :=
  depConstrOLQInsert a q.

Definition dequeueHelpOLQ (outer : A) 
  (q : OLQ) (m : option (OLQ * A)) : 
  option (OLQ * A) :=
@option_rect
  (OLQ * A)
  (fun _ => option (OLQ * A))
  (fun (p : (OLQ * A)) => Some 
    (depConstrOLQInsert outer (fst p) , 
      (snd p)))
  (Some (depConstrOLQEmpty, outer))
  m.

Definition dequeueOLQ : 
  OLQ -> option (OLQ * A) :=
  depRecOLQ (option (OLQ * A)) None dequeueHelpOLQ.
\end{lstlisting}
\end{minipage}
\begin{minipage}{0.485\textwidth}
\begin{lstlisting}[mathescape=true, language=coq]
Definition enqueueTLQ (a : A) (q : TLQ) : TLQ :=
  depConstrTLQInsert a q.

Definition dequeueHelpTLQ (outer : A) 
  (q : TLQ) (m : option (TLQ * A)) : 
  option (TLQ * A) :=
@option_rect
  (TLQ * A)
  (fun _ => option (TLQ * A))
  (fun (p : (TLQ * A)) => Some 
    (depConstrTLQInsert outer (fst p) , 
      (snd p)))
  (Some (depConstrTLQEmpty, outer))
  m.

Definition dequeueTLQ : 
  TLQ -> option (TLQ * A) :=
  depRecTLQ (option (TLQ * A)) None dequeueHelpTLQ.
\end{lstlisting}
\end{minipage}
\caption{Definitions for \lstinline{enqueue} and \lstinline{dequeue} over \lstinline{OLQ} on the left, and their repaired versions over \lstinline{TLQ} on the right.}
\label{fig:queue-fns}
\end{figure}

\begin{figure}
\centering
\begin{minipage}{0.485\textwidth}
\begin{lstlisting}[mathescape=true, language=coq]
Definition returnOrEnqOLQ (a : A) 
  (m : option (OLQ * A)) : (OLQ * A) :=
  @option_rect
    (OLQ * A)
    (fun _ => prod OLQ A)
    (fun (p : (OLQ * A)) => 
      (enqueueOLQ a (fst p), snd p))
    (depConstrOLQEmpty, a)
    m.
    
Theorem dequeueEnqueue (a : A) (q : OLQ) : 
  dequeueOLQ (enqueueOLQ a q) 
  = Some (returnOrEnqOLQ a (dequeueOLQ q)).
\end{lstlisting}
\end{minipage}
\begin{minipage}{0.485\textwidth}
\begin{lstlisting}[mathescape=true, language=coq]
Definition returnOrEnqTLQ : (a : A) 
  (m : option (TLQ * A)) : (TLQ * A) := 
  @option_rect 
    (TLQ * A)
    (fun _ => prod TLQ A)
    (fun (p : (TLQ * A)) => 
      (enqueueTLQ a (fst p), snd p))
    (depConstrTLQEmpty, a) 
    m.
     
Theorem dequeueEnqueue (a : A) (q : TLQ) : 
  eq_deq_ret
    (dequeueTLQ (enqueueTLQ a q))
    (Some 
      (returnOrEnqTLQ a (dequeueTLQ q))).
\end{lstlisting}
\end{minipage}
\caption{Main theorem relating dequeue and enqueue, stated for \lstinline{OLQ} on the left and \lstinline{TLQ} on the right. We repair this theorem's proof from \lstinline{OLQ} to \lstinline{TLQ}. Here, \lstinline{eq_deq_ret} is the equivalence relation \lstinline{eq_queue} lifted to the return type of \lstinline{dequeueTLQ}.}
\label{fig:dequeue-enqueue-theorem-statement}
\end{figure}

\subsection{Variations on a Theme of Queues}
\label{sec:case2}

Next, we repair functions and proofs across a change in implementation of a queue data structure. 
This is motivated by an example from \citet{angiuli2021},
which showed that quotient types can be used to adjust
certain relations more general than equivalences into equivalences
for use with transport in Cubical Agda. That class of changes was cited in the \textsc{Pumpkin P}i paper as an example that could not be expressed naturally in Rocq with the original framework. With our extensions to \textsc{Pumpkin P}i, we can express
this using setoids. Our Rocq implementation of this case study can be found in the artifact 
~\href{https://github.com/uwplse/pumpkin-pi/blob/oopsla2025/plugin/case-studies/two_list_queue_equivalence_repair_tool.v}{\circled{10}}.

\paragraph{Types \& Configuration} Our first implementation \lstinline{OLQ} represents queues using a single list. Elements enqueue at the front of the list and dequeue from the back of the list. This is simple, but the dequeue operation runs in linear time.
Our second implementation \lstinline{TLQ} uses a two list representation of queues. Elements enqueue at the front of the first list, and dequeue from the front of the second list, reversing the first list onto the second when the second is empty. This defines an amortized constant time dequeue operation.

Each two list queue \lstinline{(l1, l2)} corresponds to the queue \lstinline{l1 ++ (rev l2)}, but multiple two list queues correspond to a single one list queue. Thus, we use the equivalence relation 
\linebreak \lstinline{(l1, l2)}$\hspace{-0.02cm}$ $\sim$$\hspace{-0.02cm}$ \lstinline{(l3, l4) $\iff$$\hspace{-0.02cm}$ l1$\hspace{-0.04cm}$ ++$\hspace{-0.04cm}$ (rev$\hspace{-0.04cm}$ l2)$\hspace{-0.04cm}$ =$\hspace{-0.04cm}$ l3$\hspace{-0.04cm}$ ++$\hspace{-0.04cm}$ (rev$\hspace{-0.04cm}$ l4)}
and consider \lstinline{TLQ} as a setoid. These two types are setoid equivalent along the expected correspondence,
which lets us define the repair configuration. See Figure~\ref{fig:queue-types} for the types, and the artifact for the repair configuration.

\paragraph{Function Repair}
We now use our extension to repair
functions across this change. We provide the standard queue API by repairing \lstinline{enqueueOLQ} and \lstinline{dequeueOLQ}, as well as the helper functions \lstinline{dequeueHelpOLQ} and \lstinline{returnOrEnqTLQ}. The first three can be found in Figure~\ref{fig:queue-fns}, and the last in Figure~\ref{fig:dequeue-enqueue-theorem-statement}.

In the previous case study, our automation succeeded in
generating every function's properness proof.
This time, however, while the proofs that \lstinline{enqueueTLQ} and \lstinline{dequeueTLQ} are proper are generated automatically, the properness proofs for \lstinline{dequeueHelpTLQ} and \lstinline{returnOrEnqTLQ} failed to generate and needed to be supplied manually. Also, we need to define multiple equivalence relations, since the return type of \lstinline{dequeueTLQ} is \lstinline{option (TLQ * A)} and our automation does not yet automatically lift the equivalence over \lstinline{TLQ} to types including \lstinline{TLQ}. The user must provide these equivalences to \textsc{Pumpkin P}i.

\paragraph{Proof Repair}
We prove a theorem \lstinline{dequeueEnqueueOLQ}, found in Figure~\ref{fig:dequeue-enqueue-theorem-statement}, stating that enqueue and dequeue commute in the expected way, using \lstinline{returnOrEnq} as a helper function. We repair this automatically, with none of the workarounds from the previous case study, since the proof of \lstinline{dequeueEnqueueOLQ} 
is defined using the \lstinline{option_rect} rather
than \lstinline{depElimProp}. We also repair the proof of \lstinline{dequeueEmptyOLQ}, providing an algebraic specification for the repaired datatype.

\begin{figure}
\begin{lstlisting}[mathescape=true, language=coq]
Definition fastDequeueTLQ (q : TLQ) :=
  let (l1, l2) := q in match l1, l2 with
  | [] , [] => None
  | h1 :: t1 , [] => 
      Some (([] , tl (rev l1)), hd h1 (rev l1))
  | _ , h2 :: t2 => Some ((l1, t2), h2)
  end.
\end{lstlisting}
\caption{Fast dequeue function for two list queues.}
\label{fig:dequeue-fast}
\end{figure}

\begin{figure}
\centering
\begin{minipage}{0.49\textwidth}
\begin{lstlisting}[mathescape=true, language=coq]
Definition CLPoly := list nat.
Definition eq_CLPoly (l1 l2 : CLPoly) :=
  removeLeadingZeros l1 = 
    removeLeadingZeros l2.
\end{lstlisting}
\end{minipage}
\begin{minipage}{0.49\textwidth}
\begin{lstlisting}[mathescape=true, language=coq]
Definition CEPPoly := list (nat * nat).
Definition eq_CEPPoly (p1 p2 : CEPPoly) :=
  forall (exp : nat), 
    coeff p1 exp = coeff p2 exp.
\end{lstlisting}
\end{minipage}
\caption{Definitions and equivalence relations for \lstinline{CLPoly} and \lstinline{CEPPoly}. \lstinline{removeLeadingZeros l} removes leading zeros from \lstinline{l}, while \lstinline{coeff p exp} is the \lstinline{n}th degree coefficient of \lstinline{p}. We provide instances of \linebreak \lstinline{Equivalence eq_CLPoly} and \lstinline{Equivalence eq_CEPPoly} in the artifact.}
\label{fig:polynomial-types}
\end{figure}

\begin{figure*}
\begin{minipage}{.49\textwidth}
\begin{lstlisting}[mathescape=true, language=coq]
Definition depConstrCLPoly 
  (l : opaque_list) 
  (p : noLeadingZeros l) : CLPoly.
Definition depRecCLPoly (C : Type) 
  (X : forall (l : opaque_list) 
    (p : noLeadingZeros l), C) 
  (p : CLPoly) : C.
\end{lstlisting}
\end{minipage}
\begin{minipage}{.49\textwidth}
\begin{lstlisting}[mathescape=true, language=coq]
Definition depConstrCEPPoly 
  (l : list nat) 
  (p : noLeadingZeros l) : CEPPoly.
Definition depRecCEPPoly (C : Type) 
  (X : forall (l : list nat) 
    (p : noLeadingZeros l), C) 
  (p : CEPPoly) : C.
\end{lstlisting}
\end{minipage}
\caption{The types of \lstinline{depConstr} and \lstinline{depRec} for \lstinline{CLPoly} on the left and \lstinline{CEPPoly} on the right, where \lstinline{opaque_list} is an alias for \lstinline{list nat}.}
\label{fig:polynomial-config-partial}
\end{figure*}

\begin{figure*}
\begin{minipage}{.49\textwidth}
\begin{lstlisting}[mathescape=true, language=coq]
Definition addCLPoly (p1 p2 : CLPoly) := 
  depRecCLPoly CLPoly
    (fun (l : opaque_list) 
      (p : noLeadingZeros l) =>
       depRecCLPoly CLPoly
         (fun (l0 : opaque_list) 
           (p0 : noLeadingZeros l0) =>
           depConstrCLPoly (addLists l l0) 
             (addListsNoLeadingZeros l 
               l0 p p0)) 
        p2) 
    p1.

Definition evalCLPoly (p : CLPoly) 
  (n : nat) :=
  depRecCLPoly nat 
    (fun (l : opaque_list) 
      (proof : noLeadingZeros l) => 
        evalList l n) 
    p.
\end{lstlisting}
\end{minipage}
\begin{minipage}{.49\textwidth}
\begin{lstlisting}[mathescape=true, language=coq]
Definition addCEPPoly (p1 p2 : CEPPoly) := 
  depRecCEPPoly CEPPoly
    (fun (l : opaque_list) 
      (p : noLeadingZeros l) =>
       depRecCEPPoly CEPPoly
         (fun (l0 : opaque_list) 
           (p0 : noLeadingZeros l0) =>
           depConstrCEPPoly (addLists l l0) 
             (addListsNoLeadingZeros l 
               l0 p p0)) 
        p2) 
    p1.

Definition evalCEPPoly (p : CEPPoly) 
  (n : nat) :=
  depRecCEPPoly nat 
    (fun (l : opaque_list) 
      (proof : noLeadingZeros l) => 
        evalList l n) 
    p.
\end{lstlisting}
\end{minipage}
\caption{The definitions of addition and evaluation for polynomials. The original, over \lstinline{CLPoly}, is on the left, and the repaired version over \lstinline{CEPPoly} is on the right.}
\label{fig:polynomial-fns}
\end{figure*}

\begin{figure*}
\begin{minipage}{.49\textwidth}
\begin{lstlisting}[mathescape=true, language=coq]
Theorem addComm :
  forall (p1 p2 : CLPoly),
    eq_CLPoly (add p1 p2) (add p2 p1).
\end{lstlisting}
\end{minipage}
\begin{minipage}{.49\textwidth}
\begin{lstlisting}[mathescape=true, language=coq]
Theorem evalRespectsAdd :
  forall (p1 p2 : CLPoly) (n : nat),
    eval (add p1 p2) n = 
      (eval p1 n) + (eval p2 n).
\end{lstlisting}
\end{minipage}
\caption{The types of \lstinline{addComm} and \lstinline{evalRespectsAdd} for \lstinline{CLPoly}. These theorems were repaired to \lstinline{CEPPoly}.}
\label{fig:polynomial-theorem-statements}
\end{figure*}

\paragraph{Further Steps}
As in the previous case study, we have finished repairing proofs,
but our repaired implementation of dequeue is inefficient.
Thus, we implement the fast version of dequeue described earlier, found in Figure~\ref{fig:dequeue-fast}, and can follow the same methodology as in the previous case study to port repaired proofs
to use this fast version. We first show pointwise equality of the functions, and then rewrite inside of our proofs, giving us repaired proofs about the fast dequeue function.

\subsection{Polynomial Polynomials}
\label{sec:case3}
For our third case study, we provide sparse and dense representations of univariate polynomials with natural number coefficients, both of which are represented using setoids. The former representation uses the simplest possible data type to represent a polynomial, while the latter representation will mimic the way polynomials are commonly written, which will utilize far less storage for polynomials of high degree with many terms having a coefficient of 0.
We implement and repair addition of polynomials and evaluation of polynomials on a natural number, as well as proofs that addition is commutative and that evaluation respects addition. Our proofs can be found in the artifact 
~\href{https://github.com/uwplse/pumpkin-pi/blob/oopsla2025/plugin/case-studies/polynomial.v}{\circled{11}}.

\paragraph{Types \& Configuration}
Our first representation of polynomials, \lstinline{CLPoly} (short for coefficient list polynomial), is as lists of natural numbers. The members of the list are the coefficients of the polynomial in order of decreasing degree. For example, $x^2 + 3$ is represented as \lstinline{[1; 0; 3]}. Two \lstinline{CLPoly}s are equivalent if they are equal after removing leading zeros. Our second representation, \lstinline{CEPPoly} (short for coefficient-exponent pair polynomial), is lists of pairs of natural numbers. Each pair \lstinline{(c, exp)} represents a monomial in a sum, with \lstinline{c} the coefficient and \lstinline{exp} the exponent of the monomial, with duplicate exponents allowed. Using the same example, $x^2 + 3$ would be represented as \lstinline{[(1, 2); (3, 0)]}. Two members of \lstinline{CEPPoly} are equivalent if the polynomial they represent has the same coefficients. Both representations are found in Figure~\ref{fig:polynomial-types}.

Notice that every equivalence class in either of the setoids uniquely defines exactly one polynomial. Thus, the isomorphism between these setoids maps each member of the class representing polynomial $p$ in one setoid to a member of the class representing polynomial $p$ in the other setoid. 

Next, we define a configuration for repair.
 \textsc{Pumpkin P}i's configurations are based on decomposing an equivalence between inductive types,
 but here, both types are setoids.
 We choose an inductive type equivalent to both of our setoids, and the configuration components for our setoids have the structure of that type. For this, we choose the type of canonical representatives of \lstinline{CLPoly}, the sigma type:

 \begin{lstlisting}[language=coq]
   {l : list nat | noLeadingZeros l}
 \end{lstlisting}
 
 The types of \lstinline{depConstr} for both \lstinline{CLPoly} and \lstinline{CEPPoly} take the shape of the constructor of this type, as do the types of the eliminators and $\iota$-reduction rules. We show this for part of our configuration for \lstinline{CLPoly} and \lstinline{CEPPoly} in Figure~\ref{fig:polynomial-config-partial}. 

\paragraph{Function Repair}
We use \lstinline{CLPoly} as our source setoid and \lstinline{CEPPoly} as our target setoid. 
We define \lstinline{add} and \lstinline{eval}, representing addition and evaluation, explicitly annotated with the components of the configuration.
We automatically repair these functions using our extension
to \textsc{Pumpkin P}i. The original definitions and the repaired versions can be found in Figure~\ref{fig:polynomial-fns}. Our extension automatically
proves that \lstinline{eval} is proper, but fails to show \lstinline{add} is proper, so we do so manually.

\paragraph{Proof Repair}
We repair proofs of \lstinline{addComm}, which states that \lstinline{add} is commutative, and \linebreak \lstinline{evalRespectsAdd}, which states that evaluation distributes over addition. The statements of these theorems can be found in Figure~\ref{fig:polynomial-theorem-statements}. 
Both use \lstinline{depElimProp}, so we reconfigure \textsc{Pumpkin P}i to use specialized versions of \lstinline{depElimProp}, as we did in the first case study. From there, both proofs repair automatically. Neither of these proofs use rewriting. Thus, to demonstrate lifting setoid rewrites to setoid rewrites, we also repair three simple proofs about polynomials which specifically use setoid rewriting.

\paragraph{Further Steps}
The constructor for \lstinline{CLPoly}, seen in Figure~\ref{fig:polynomial-config-partial}, accepts an argument of type \newline \lstinline{opaque_list}. Because \lstinline{CLPoly} is internally \lstinline{list nat}, \textsc{Pumpkin P}i will attempt to repair \emph{all} instances of \lstinline{list nat}. However, we define functions over \lstinline{list nat} which we do not want to repair. Thus, we define an alias \lstinline{opaque_list} for \lstinline{list nat} and tell \textsc{Pumpkin P}i not to repair \lstinline{opaque_list}, as well as any other functions and theorems that should not be repaired.
This behavior and our strategy for dealing with it
are inherited from \textsc{Pumpkin P}i. 

\section{Correctness}
\label{sec:discussion}

Up to now, we have been imprecise with what it means for repair to be conducted correctly. The original \textsc{Pumpkin P}i paper \cite{ringer2021pldi} 
outlines a definition for correctness of repair using univalent type theory. We review that definition here, giving the needed background in univalent type theory. Then, we implement this definition in a proof assistant which has the necessary univalent type theory.
This allows us to, for the first time, construct proofs that functions and theorems were correctly repaired.

To state and prove correctness of repair, we need a notion of heterogeneous equality. The \textsc{Pumpkin P}i paper uses \textit{dependent path equality} in a univalent type system for this purpose. Univalence states that equivalence is equivalent to equality~\cite{hottbook, CCHM, angiuli2017}: any equivalence of types corresponds to a unique proof of equality between those two types. We term these equalities \textit{path equalities}. Given a path equality \lstinline[mathescape=true]{p : A $\equiv$ B} between types and given elements \lstinline{a : A}, \lstinline{b : B} of those types, we can construct the type of dependent path equalities between those elements, notated \lstinline{PathP p a b}. This type is inhabited if and only if the equivalence of types corresponding to \lstinline{p} by univalence maps \lstinline[mathescape=true]{a} to \lstinline[mathescape=true]{b}.

The repair transformation is parametrized by an equivalence \lstinline[mathescape=true]{f : A $\to$ B} between the old and new types. By univalence, this equivalence gives an equality between those types. This equality of types in turn extends to equalities of types built using those types: for instance, if \lstinline[mathescape=true]{A $\equiv$ B}, then \lstinline[mathescape=true]{A $\to$ C $\equiv$ B $\to$ C}. For any repaired term, then, there is an equality between the type of the original term and the type of the repaired term. We obtain this equality inductively in much the same way we do repair inductively. Whenever an inductive type appears in the old term, if that type is \lstinline[mathescape=true]{A}, we use the equality proof corresponding to \lstinline[mathescape=true]{f}. Whenever another inductive type appears, we use \lstinline{refl} as the equality proof. When the other rules would apply, we have theorems constructing an equality proof from the proofs for the constituent terms. Thus, we can construct the type of dependent path equalities between these terms. If that type is inhabited, we say that the term was repaired correctly. 

Thus, to use this approach to construct proofs of correct repair, we need to work in a type system that supports this notion of dependent path equality. For this, we choose Cubical Agda.
It is true that we could have just as well used Rocq's HoTT library~\cite{bauer2017coqhott}, which treats univalence as an axiom. However, Rocq's HoTT library also fundamentally changes how the \lstinline{Prop} universe works. Switching to such a library for the sake of constructing correctness proofs would have imposed as much overhead for us as directly using a different proof assistant that has univalence, and may have been extra confusing to readers as it would have looked the same as Rocq while relying on fundamentally different dependencies that change the type theory.
Therefore, we instead switch to Cubical Agda. We also like the bonus that terms in Cubical Agda compute, since Cubical Agda has univalence as a theorem rather than an axiom.

%b Using HoTT in Coq takes a bit more than just adding an axiom, due to the way `Prop` works in Coq, so it would really amount to using an existing Coq HoTT library. We could have done this, but it would have had as much overhead for us as just using a different proof assistant that has univalence, and might have even been more confusing since it would have looked the same while relying on fundamentally different dependencies that change the type theory. We also like the bonus that the terms in Cubical Agda compute rather than relying on an axiom.

As a technical detail, we restrict ourselves to working with types which are \textit{h-sets}: that is, types where uniqueness of identity proofs holds. As a bonus, Cubical Agda allows a direct implementation of quotient types via the following higher inductive type:
\begin{lstlisting}
  data _/_ (A : Type) (R : A $\rightarrow$ A $\rightarrow$ Type) : Type
    [_] : (a : A) $\rightarrow$ A / R
    eq/: (a$_1$ a$_2$: A) $\rightarrow$ (r: R a$_1$ a$_2$) $\rightarrow$ [ a$_1$ ] $\equiv$ [ a$_2$ ]  
    squash/ : (x y : A / R) $\rightarrow$ (p q : x $\equiv$ y) $\rightarrow$ p $\equiv$ q 
\end{lstlisting}
The first constructor is the constructor for an element of a quotient type as described in Section~\ref{sec:setoids}, and the second encodes the equality property for quotient types. The third constructor enforces that quotient types are h-sets. We will use this type instead of setoids when specifying correctness of repair in Cubical Agda. 
As a result, the additional rules to repair equivalence relations from Figure~\ref{fig:quotient-transform} in Section~\ref{sec:approach-coq} are not needed when conducting repair in Cubical Agda. The transformation from Figure~\ref{fig:pumpkin-pi-transformation} in Section~\ref{sec:approach-coq} works, assuming the same annotations as in Rocq, noting that we cannot reuse any of \textsc{Pumpkin P}i's automation in Cubical Agda. There is only one additional restriction: for a quotient type \lstinline{Q}, every motive \lstinline{P : Q $\to$ Set} comes with the requirement that \linebreak \lstinline{((x : Q)  $\to$ isSet (P x))}, to ensure that we actually do stay in the h-set fragment of Cubical Agda. 
To demonstrate that our repair methodology still works under this paradigm, we manually followed the transformation to repair the
functions and proofs from the first two case studies in Section~\ref{sec:overview}, which can be found in the artifact 
~\href{https://github.com/uwplse/pumpkin-pi/blob/oopsla2025/plugin/case-studies/grothendieck_int_equiv.agda}{\circled{12}} 
~\href{https://github.com/uwplse/pumpkin-pi/blob/oopsla2025/plugin/case-studies/equivalence_queue.agda}{\circled{13}}.

\begin{figure*}
\begin{lstlisting}[language=cubicalagda]
  lamOK: {T} {F}
    (f: (t: T i0) $\to$ F i0 t) (f': (t: T i1)$\to$ F i1 t)
    (b$\equiv$b' : $\forall$ {t : T i0} {t' : T i1} 
      (t$\equiv$t' : PathP ($\lambda$ i $\to$ T i) t t') $\to$
      PathP ($\lambda$ i $\to$ F i (t$\equiv$t' i)) (f t) (f' t')) $\to$
    PathP ($\lambda$ i $\to$ $\forall$ (t : T i) $\to$ F i t) f f'
  lamOK {T} {F} f f' b$\equiv$b' = funExtDep b$\equiv$b'
\end{lstlisting}
\caption{A theorem showing that the \textsc{Lam} rule is correct. Here, \lstinline{i}, \lstinline{i0}, and \lstinline{i1}
are terms of the interval type, which is a primitive construct in cubical used to define path equalities.
The rest is analogous to Figure~\ref{fig:pumpkin-pi-transformation}: \lstinline{f} is the left function in the rule, \lstinline{f'} is the right function, \lstinline{F i0} is the type of \lstinline{f}, \lstinline{F i1} is the type of \lstinline{f'}, and all other subterms have the same names.}
\label{fig:lamOK}
\end{figure*}

Then, we go about proving theorems which state that each repair rule repairs terms correctly, given that the inputs to the rule are themselves correctly repaired. Some of these theorems are generic across all types. For example, we internally prove correctness of the \textsc{Lam} rule of the transformation from Figure~\ref{fig:pumpkin-pi-transformation} in Figure~\ref{fig:lamOK}, which is generic across any repair instance. 

\begin{figure*}
\begin{lstlisting}[mathescape=true,language=cubicalagda]
elimOK : 
  $\forall$ (a : $\mathbb{N}$) (b : Int / rInt) (a$\equiv$b : PathP ($\lambda$ i $\to$ $\mathbb{N}$$\equiv$Int/rInt i) a b) $\to$
  $\forall$ (PA : $\mathbb{N}$ $\to$ Type) (PB : Int / rInt $\to$ Type) (PBSet : $\forall$ b $\to$ isSet (PB b)) $\to$
  $\forall$ (PA$\equiv$PB : 
    $\forall$ a b (a$\equiv$b : PathP ($\lambda$ i $\to$ $\mathbb{N}$$\equiv$Int/rInt i) a b) $\to$ 
    PathP ($\lambda$ i $\to$ Type) (PA a) (PB b)) $\to$
  $\forall$ (PAO : PA zero) (PBO : PB depConstrInt/rInt0) $\to$
  $\forall$ (PAO$\equiv$PBO : PathP ($\lambda$ i $\to$ PA$\equiv$PB zero depConstrInt/rInt0 depConstr0OK i) PAO PBO) $\to$
  $\forall$ (PAS$\hspace{-0.1cm}$ :$\hspace{-0.1cm}$ $\forall$ a $\to$ PA a $\to$ PA (suc a)) (PBS$\hspace{-0.1cm}$ :$\hspace{-0.1cm}$ $\forall$ b $\to$ PB b $\to$ PB (depConstrInt/rIntS b)) $\to$
  $\forall$ (PAS$\equiv$PBS :
    $\forall$ a b (IHa$\hspace{-0.11cm}$ :$\hspace{-0.11cm}$ PA a) (IHb$\hspace{-0.1cm}$ :$\hspace{-0.11cm}$ PB b) a$\equiv$b (IHa$\equiv$IHb$\hspace{-0.11cm}$ :$\hspace{-0.11cm}$ PathP ($\lambda$ i$\hspace{-0.11cm}$ $\to$$\hspace{-0.11cm}$ PA$\equiv$PB a b a$\equiv$b i) IHa IHb)$\hspace{-0.11cm}$ $\to$$\hspace{-0.11cm}$
    PathP ($\lambda$ i $\to$ PA$\equiv$PB (suc a) (depConstrInt/rIntS b) (depConstrSOK a b a$\equiv$b) i)
      (PAS a IHa)
      (PBS b IHb)) $\to$
  PathP ($\lambda$ i $\to$ PA$\equiv$PB a b a$\equiv$b i)
    (Nat.elim {A = PA} PAO PAS a)
    (depElimSetInt/rInt PB PBSet PBO PBS b)
\end{lstlisting}
\caption{The theorem stating the correctness condition for the repaired dependent eliminator for a simple example type, which has been proven internally in Cubical Agda. 
This theorem shows that, if all the inputs to the eliminator correspond to each other across the isomorphism, then the output of the eliminator applications also corresponds across that isomorphism.
Here, \lstinline{depConstr0OK} and \lstinline{depConstrSOK} are the correctness proofs of the repaired constructors, also proven internally.}
\label{fig:elimOK}
\end{figure*}

Other rules are stated specifying the types being repaired. For example,
we proved the repaired eliminator we defined for a simple quotient equivalence was correct. Our source type was $\mathbb{N}$, and our target was \lstinline{Int / rInt}, where \lstinline{Int = $\mathbb{N} \uplus \mathbb{N}$} and \lstinline{rInt} relates \lstinline{inl n} and \lstinline{inr n}. The dependent constructors and eliminators correspond to the usual ones for $\mathbb{N}$. 
The correctness condition for a repaired
eliminator for a given configuration was stated externally in the \textsc{Pumpkin P}i paper, but it was not proven for any type.
We adapted this theorem to Cubical Agda for our example and, for the first time,
proved that it held. The type of the rule showing that the dependent eliminator repairs correctly can be found in Figure~\ref{fig:elimOK}.

Using these rules, we were able to compose the correctness proofs to show the correctness
of repaired \emph{functions}, like addition:
\begin{lstlisting}[language=cubicalagda]
  addCorrect : $\forall$ (a b : $\mathbb{N}$) (a' b' : Int / rInt) $\to$
    $\forall$$\hspace{-0.04cm}$ (pa$\hspace{-0.04cm}$ :$\hspace{-0.04cm}$ PathP$\hspace{-0.04cm}$ ($\lambda$$\hspace{-0.04cm}$ i$\hspace{-0.04cm}$ $\to$$\hspace{-0.04cm}$ Nat$\equiv$Int/rInt$\hspace{-0.04cm}$ i)$\hspace{-0.04cm}$ a$\hspace{-0.02cm}$ a')$\hspace{-0.04cm}$ (pb$\hspace{-0.04cm}$ :$\hspace{-0.04cm}$ PathP$\hspace{-0.04cm}$ ($\lambda$$\hspace{-0.04cm}$ i$\hspace{-0.04cm}$ $\to$$\hspace{-0.04cm}$ Nat$\equiv$Int/rInt$\hspace{-0.04cm}$ i)$\hspace{-0.04cm}$ b$\hspace{-0.04cm}$ b')$\hspace{-0.04cm}$ $\to$
    $\phantom{\forall}$ PathP ($\lambda$ i $\to$ Nat$\equiv$Int/rInt i) (add' a b) (addInt/rInt' a' b')
\end{lstlisting}

We also were able to prove theorems about these functions. For instance, below is the type of a term proving that our proof that addition is commutative on $\mathbb{N}$, \lstinline{addCommNat}, was correctly repaired to a proof that addition is commutative on our new type:

\begin{lstlisting}[mathescape=true,language=cubicalagda]
addCommCorrect :
  (a : $\mathbb{N}$) (a' : Int / rInt) (pa : PathP ($\lambda$ i $\to$ Nat$\equiv$Int/rInt i) a a') $\to$
  (b : $\mathbb{N}$) (b' : Int / rInt) (pb : PathP ($\lambda$ i $\to$ Nat$\equiv$Int/rInt i) b b') $\to$
  PathP ($\lambda$ i $\to$ addCorrect a b a' b' pa pb i $\equiv$ addCorrect b a b' a' pb pa i) 
    (addCommNat a b) (addCommInt/rInt a' b')
\end{lstlisting}

When constructing proofs of correct repair for theorems, because of difficulties when composing different \lstinline{PathP}s, our rules for proving correct repair do not always compose directly in their current formulation. As a result, if Cubical Agda had the facilities for building automation, we would not currently have a complete procedure for constructing proofs of correct repair fully automatically. However, we are able to prove these theorems directly in such cases, demonstrating that our repair procedure produces the desired output. All of the rules we prove, as well as the proofs of correct repair in our example, can be found in the artifact
~\href{https://github.com/uwplse/pumpkin-pi/blob/oopsla2025/plugin/case-studies/equivalence_int_abs.agda}{\circled{14}}.

\section{Related Work}
\label{sec:related}

\hspace{\parindent}\textbf{Proof Repair.}
This work extends the \textsc{Pumpkin P}i~\cite{ringer2021pldi} proof repair 
transformation and Rocq plugin
to support quotient type equivalences,
a class of changes previously not supported.
\textsc{Pumpkin P}i has some more mature automation
for other classes of changes, like automatic
search for configurations, that we do not yet extend
to work for quotient type equivalences. 
Proof repair was first introduced in parallel
by \citet{ringer2018cpp} and \citet{robert2018},
with strong influence from program repair~\cite{Monperrus2015}.
\textsc{Sisyphus}~\cite{gopinathan2023} is a recent proof repair tool that, like our work, can handle changes in behavior (using a mix of
dynamic and static techniques). However, \textsc{Sisyphus} repairs proofs of imperative OCaml programs verified in Rocq using an embedded separation logic, whereas our work repairs proofs that are written in Rocq directly.

\textbf{Univalent Foundations.} Parts of this project are grounded in Cubical Agda, and parts assume a univalent metatheory. 
Cubical Agda is an implementation of cubical type theory~\cite{vezzosi2019ogcubicalagda}. Cubical type theory~\cite{CCHM, CHM, angiuli2017} was developed to give a constructive account of the univalence axiom. 
When working in Cubical Agda, we are able to state and prove internal correctness of parts of our repair transformation and have a computational interpretation of functional extensionality. Cubical type theory itself is a derivative of Voevodsky's homotopy type theory~\cite{hottbook}, which presents the univalence axiom non-constructively. Homotopy type theory has additionally been implemented in Rocq as the HoTT library~\cite{bauer2017coqhott}. Univalence leads to a related idea, the Structure Identity Principle, which states that the type of equalities of structures is equivalent to the type of isomorphisms of those structures \cite{hottbook, coquandIsoIsEq, ahrens2022univalenceprinciple}.

\textbf{Proof Reuse and Transfer.}
Proof repair is an instance of proof reuse, which seeks to use existing proofs in new goals.  
Other work in proof reuse includes CoqEAL~\cite{refinementsforfree} which uses refinement relations to verify properties of efficient functions using proofs on functions that are easy to reason about.
CoqEAL can handle relations more general than equivalences, but does not include support
for porting proofs across those changes.
In Isabelle/HOL, the Transfer package~\cite{isabelletransfer} uses automation to transfer proofs between types. Both approaches require the source and target type to remain in the codebase, unlike proof repair.
A complementary approach is to design proofs to 
be more reusable or more robust to changes from the start~\cite{Woos2016, cpdt, Delaware2011}.
More work on proof reuse can be found in the QED at Large~\cite{PGL-045} survey of proof engineering.

Work has been done to implement transfer tools in Rocq that approximate or
externally implement automation corresponding to univalent transport.
\citet{equivalencesforfree} defines univalent parametricity, which allows transport of a restricted class of functions and theorems. 
Univalent parametricity implements an ad hoc form
of transport that only sometimes requires
functional extensionality, and 
in many cases is axiom-free.
It also includes a form of type-directed search
to transport terms by way of type classes,
something that proof repair tools like
\textsc{Pumpkin P}i and our extension still lack.
Subsequent work introduces a white-box transformation~\cite{marriage} similar to the repair transformation from \textsc{Pumpkin P}i, which \citet{ringer2021proof} describes as developed in parallel with mutual influence.
None of these support quotient type equivalences
like our work does, though it is possible that
by leaning further on functional extensionality,
one could use these tools with quotient
type equivalences.

More recent work called, Trocq~\cite{cohen2024trocq} implements external transfer for Rocq that
directly supports relations more general than equivalences, like CoqEAL, but also supports proofs.
Like \textsc{Pumpkin P}i, Trocq goes out of its way to avoid depending on axioms like univalence
and functional extensionality.
Trocq's motivation of supporting transfer of proofs across relations more general than equivalences
is similar to the motivation of our extensions to \textsc{Pumpkin P}i, with two differences:
(1) our work supports a more limited class of relations that can be described as equivalences between quotient types, and
(2) for that class of changes, by extending \textsc{Pumpkin P}i's proof term transformation,
our work makes it possible to remove the old version of a type 
after applying repair.
The major benefit of our tool relative to Trocq comes from (2)---while all
proof repair tools implement a kind of transfer, not all transfer methods implement repair,
and Trocq does not implement repair.

\textbf{Quotients and Equivalences.} Our work uses quotient types \cite{hofmann1995extensional} to expand the scope of proof repair. \citet{maietti1998effective} shows that, in intensional Martin-L\"{o}f type theory with uniqueness of identity proofs and at least two universes, the existence of effective quotients implies the law of the excluded middle for types in the first universe. Quotient types exist in other proofs assistants besides Cubical Agda, like Isabelle/HOL~\cite{isabelle, isabelletool}, as well as Lean~\cite{lean} by way of axioms. 
\citet{bortin2010isabellequotient} use quotient types to construct theories in Isabelle, like multisets and finite sets as quotients of lists.
Rocq does not have quotient types, but it does have setoids~\cite{setoid}, which do not explicitly form equivalence classes like quotients do.
Setoid type theory uses a setoid model to justify
the axioms needed to represent quotient types~\cite{setoidtt}. XTT uses Bishop sets, analogous to setoids, to construct a type theory where all types have definitional uniqueness of identity proofs \cite{xtt}.
We draw on quotient types for our
work in Cubical Agda, and we draw on setoids
for our work in Rocq.

Our idea for extending proof repair using quotient type
equivalences to begin with comes from
\citet{angiuli2021}, which shows that 
certain relations more general than equivalences
can be represented this way.
The first example present in that paper is the queue example which we have also studied in our work.
Because that work uses transport, it requires the user to keep both versions of the type in their codebase. We avoid that problem, but also have to reason more closely about the inductive structure of our types.
In doing so, we extend proof repair to support a new
class of changes described as missing from the original \textsc{Pumpkin P}i work~\cite{ringer2021pldi}.

% not just my stuff now, also Ilya's

% Proof repair citations:

% - ``Proof repair across type equivalences" \cite{ringer2021pldi}

% - ``Adapting proof automation to adapt proofs" \cite{ringer2018cpp}

% - Ilya's ``Mostly Automated Proof Repair for Verified Libraries" \cite{gopinathan2023} \outline{Max note: the Sisyphus algo presented in the paper seems like it could be adapted to infer which iota to use and when to use it. Probably worth mentioning a combined approach, which would fit in well with the ``best of both worlds" framing of the paper. Also note that we can take advantage of cubical agda for internal correctness, quotient types, etc.
% }

% \paragraph*{Univalent Transport}
% \outline{marriage of univalence and parametricity~\cite{marriage}/equivalences for free. carlo paper. the two flavors of cubical.}

% Citations:

% - Hott book \cite{hottbook}

% - Coq HOTT \cite{bauer2017coqhott}

% \paragraph*{Quotient Types}
% \outline{other type theories and logics with quotients. use of quotients to implement transfer in Isabelle/HOL. ways of pretending we have quotients in Coq, and implications for future work for us.}

% Citations:

% - quotient types for representation independence \cite{angiuli2021} 

% - using quotient types in Isabelle/HOL to write general proofs and derive different implementations: \cite{bortin2010isabellequotient}
% \outline{max: this idea could be another interesting use case/new proof workflow}

\section{Conclusions \& Future Work}
\label{sec:conclusions}

We extended \textsc{Pumpkin P}i to support changes
represented by quotient type equivalences, enabling repair in situations previously untenable.
The key challenge we overcame was supporting quotient types in a proof repair algorithm built for a type theory that does not have quotient types to begin with.
We addressed this by representing quotient types using setoids, 
extending the \textsc{Pumpkin P}i algorithm and implementation to 
repair proofs about equivalence relations, and adding
new automation to dispatch newly generated proof
obligations. Our extension demonstrated success on three case studies
not supported by the original \textsc{Pumpkin P}i.
We also constructed the first internal correctness proofs for repair.
We wish to continue to improve our extension's automation and usability, and we hope to look at other kinds of types 
and relations that can be expressed even when the type theory lacks them,
as quotient types can be by way of setoids.
We also hope to tackle automation for proof repair directly in Cubical Agda,
which is challenging because Cubical Agda lacks tools for building automation (except by reflection).
We hope this will open the door to proof repair
for more sophisticated classes of changes. 

\iffalse
\section*{Acknowledgements}
We thank Amélia Liao, Reed Mullanix, Tom Jack, and the
Cubical Agda Discord server for their help in using Cubical
Agda. We thank also Carlo Angiuli for their early thoughts
on this project. This research was developed with funding
from the Defense Advanced Research Projects Agency. The
views, opinions, and/or findings expressed are those of the
author(s) and should not be interpreted as representing the
official views or policies of the Department of Defense or
the U.S. Government.
\fi
\section*{Data-Availabilty Statement}
\label{sec:das}

This paper comes with an artifact hosted on \href{https://doi.org/10.5281/zenodo.16921959}{Zenodo} containing the extension to \textsc{Pumpkin P}i we wrote, as well as all the case studies described within the paper \cite{artifact}. Instructions for reproducing the results of the artifact can be found in \lstinline{Overview.md} in the artifact. The source code for our extension to \textsc{Pumpkin P}i and our case studies can be found on \href{https://github.com/uwplse/pumpkin-pi/releases/tag/oopsla2025}{Github}. Links to specific files in the Github source have been provided where relevant throughout the paper. These links are given as circled numbers, like \circled{1}.

\section*{Acknowledgements}
% TODO thank people also
We thank Amélia Liao, Reed Mullanix, Tom Jack, and the Cubical Agda Discord server for their help in using Cubical Agda. We thank also Carlo Angiuli for their early thoughts on this project.
This material is based upon work supported by the Air Force Research Laboratory (AFRL),
the Defense Advanced Research Projects Agency (DARPA), and the Naval Information Warfare Center Pacific (NIWC Pacific) under Contracts No. N66001-21-C-4023
and FA8750-24-C-B044. 
Any opinions, findings, and conclusions or recommendations expressed in this material are those of the authors and do not necessarily reflect the views of the AFRL, DARPA, and NIWC Pacific.

\bibliographystyle{ACM-Reference-Format}
\bibliography{paper}

\end{document}